\documentclass[amsmath,amssymb,twocolumn,
 aps,prx,nofootinbib]{revtex4-2}

\usepackage{graphicx}
\usepackage{dcolumn}
\usepackage{bm}
\usepackage{xr-hyper} 
\usepackage{hyperref}
\usepackage[mathlines]{lineno}

\usepackage{mathtools}
\usepackage{amsmath}
\usepackage{amssymb}

\usepackage{comment}
\usepackage{xcolor}

\usepackage{multirow}
\begin{document}
\title{Novel Kuramoto model with inhibition dynamics modeling scale-free avalanches and synchronization in neuronal cultures}

\author{D. Lucente$^{1}$}
\altaffiliation[]{These authors have contributed equally to this study.}

\author{L. Cerutti$^{2,3}$}
\altaffiliation[]{These authors have contributed equally to this study.}

\author{M. Brofiga$^{2,3}$}  \author{A. Sarracino$^{4}$} \author{G. Parodi$^{2}$} \author{S. Martinoia$^{2,5,6}$} \author{P. Massobrio$^{2,7,\ddagger}$}\altaffiliation[]{These authors jointly supervised the study.}

\author{L. de Arcangelis$^{1,\dagger}$}
\email[Correspondence:]{Paolo.Massobrio@unige.it\\ lucilla.dearcangelis@unicampania.it}
\affiliation{
$^1$ Department of Mathematics \& Physics - University of Campania "Luigi Vanvitelli" - Caserta, Italy \\
$^2$  Department of Informatics, Bioengineering, Robotics, and Systems Engineering (DIBRIS), University of Genova, Genova, Italy \\
$^3$  Neurofacility, Istituto Italiano di Tecnologia (IIT), 16100 Genova, Italy\\
$^4$ Department of Engineering - University of Campania "Luigi Vanvitelli" - Aversa, Italy \\
$^5$ RCCS Ospedale Policlinico San Martino, Largo Rosanna Benzi 10, 16132, Genova, Italy\\
$^6$ Inter-University Center for the Promotion of the 3Rs Principles in Teaching \& Research (Centro 3R), Genova, Italy \\
$^7$ National Institute for Nuclear Physics (INFN), Genova, Italy
	}

\date{\today}

	\begin{abstract}
	Neuronal cultures exhibit a complex activity, bursts, or avalanches, characterized by the coexistence of scale \textcolor{black}{invariance} and synchronization, quite stable with the percentage of inhibitory neurons. While this bistable behavior has been already observed in the past, the characterization of the statistical properties of avalanche activity and their temporal organization is still lacking, as well as a model able to reproduce these dynamics. Here, we analyze experimental data of human neuronal cultures with controlled percentage of inhibitory neurons and characterize their statistical properties and dynamical organization. In order to model the experimental data, we propose a novel version of the Kuramoto model for two populations of oscillators, excitatory and inhibitory, implementing successfully the inhibition dynamics. The model can fully reproduce the experimental results, confirming the existence of correlations in the temporal organization of avalanche activity and the presence of an amplification - attenuation regime, as found in the human brain.
	\end{abstract}
	\maketitle

\section{Introduction}

Neuronal networks, both in vivo and in vitro, often exhibit complex activity patterns characterized by intermittent bursts of activity, named avalanches, separated by periods of relative quiescence. 
Previous studies on cortical murine cultures have shown that such systems may exhibit neuronal avalanches, with a broad range of sizes and durations, typically following power-law distributions~\cite{Beggs2003,Pasquale2008}. This dynamical regime has been proposed as a hallmark of criticality, suggesting that the brain might operate near a critical point to optimize information processing~\cite{Beggs2012}. In parallel, in vitro networks also revealed strong synchronization phenomena~\cite{gutnick1989synchronized}. The coexistence of distinct dynamical regimes, i.e. quiescent/weakly active states and  strongly synchronized states, can be explained by the concept of {\it self-organized bistability}~\cite{PhysRevLett.116.240601}. Within this framework, criticality and synchronization in excitable networks coexist~\cite{diSanto2018,buendia2021hybrid}. A minimal model capturing these features is the active rotator model~\cite{buendia2021hybrid,Buenda2022}, an extension of the Kuramoto model, which has long been a cornerstone to describe synchronization phenomena and collective dynamics~\cite{kuramoto1975international,acebron2005kuramoto}. 
Unfortunately, the original Kuramoto model is not suitable for reproducing avalanche-like or intermittent dynamics~\cite{buendia2021hybrid}. Conversely, oscillators in the active rotator model exhibit excitable behavior~\cite{shinomoto1986phase,sakaguchi1986soluble,sakaguchi1988phase}, leading the system activity to organize into avalanches whose size and duration distributions exhibit bimodality, consisting of a power-law regime followed by a bump associated with synchronized bursting~\cite{PhysRevResearch.2.013318,buendia2021hybrid}. More recently, this model has been also extended to analyze synchronization on fruitfly and human connectomes~\cite{dor2023,Buenda2022}. 

In this work, we extend the model to include an inhibitory population. Two-population Kuramoto models have been studied extensively in the context of coupled oscillators~\cite{shinomoto1986cooperative,montbrio2004synchronization,montbrio2015macroscopic,Montbri2018}, but inhibition is often modeled by simply changing the sign of the interaction constant, which only partially reflects biological interaction since it might enhance the overall activity of the network~\cite{Shimokawa2006,Montbri2018}. Here, instead, the inhibition is taken into account by introducing a time-dependence in the amplitude of the potential term entering the oscillator dynamics, ensuring that the interaction between the two populations leads to biologically plausible inhibition dynamics. Building on this biologically grounded Kuramoto model, we then directly compare experimental and numerical avalanche dynamics obtained from human induced pluripotent stem cells (hiPSCs)-derived cortical networks coupled to Micro-Electrode Arrays (MEAs). These cultures \textcolor{black}{are} systematically plated controlling the ratio between excitatory and inhibitory cells to explore how the excitation–inhibition (E:I) balance shapes network activity~\cite{Parodi2023}. In particular, we consider two E:I ratios: the physiological condition with 75$\%$ excitatory to 25$\%$ inhibitory neurons (75E:25I) and the limit case of fully excitatory networks (100E). It was recently proved that acting on the E:I balance, a modulation of the bursting activity is possible: if in vitro physiological cortical networks display a bursting-dominated activity, purely inhibitory networks abolish bursting activity, while purely excitatory assemblies display only few bursts~\cite{Parodi2023}.

Traditional analyses of neuronal avalanches have mainly focused on static properties, such as size and duration distributions and their scaling relations~\cite{Beggs2003}.
While informative, these measures do not fully capture the temporal organization of activity. To address this, we complement them with dynamical metrics, including inter-event interval distributions and conditional probabilities linking avalanche sizes to preceding inter-event times~\cite{Lombardi2016,lombardi2023beyond}. This approach provides deeper insights into the network activity and reveals different dynamical regimes.

The structure of the paper is as follows. Sec.~\ref{sec:Material_Method} describes both the experiments and the model,
including plating procedures, electrophysiological recordings, and the definition of avalanches. We also present the active Kuramoto model with our novel implementation of inhibition dynamics. Sec.~\ref{sec:Results} is devoted to the core analysis of the present work, that is the comparison between experimental and numerical results, examining both static and temporal properties of avalanche activity. We first extract size and duration distributions, showing also the verification of the scaling relation for the
conditional expectation of the avalanche size given its duration, predicted by Sethna scaling theory~\cite{dahmen1996hysteresis,perkovic1995avalanches,Kuntz2000}. Next, we discuss the properties of the inter-time distributions and explain how we define the physical
time-scale of the model through a comparison to the experimental data. We then analyze conditional
probabilities of the increments between subsequent avalanches as a function of the time elapsed between them. Finally, we discuss the implications of our findings in Sec.~\ref{sec:discussion}.

\section{Materials and methods}\label{sec:Material_Method}
\subsection{Generation and culturing of hiPSC-derived neuronal networks}\label{sec:Exp_Description}
\begin{figure*}[ht!]
    \centering
    \includegraphics[width=1.0\textwidth]{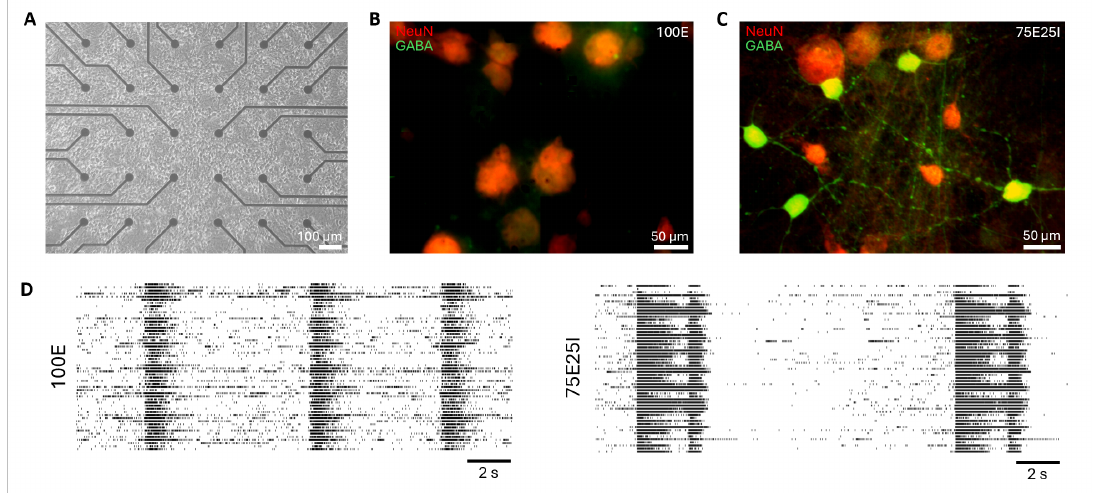}

    \caption{Representative images of hiPSC-derived neuronal cultures. A) Neuronal network plated on a micro-electrode array (MEA) at a density of $1200$ cells/$\text{mm}^2$; electrodes are visible beneath
the neuronal layer. B-C) Representative network images at DIV70 of (B) 100E and (C) 75E25I
cultures. Neurons were labelled for NeuN (red) and GABA (green) to visualize the neuronal
population and distinguish GABAergic from glutamatergic populations. D) Raster plots of the spontaneous activity of
representative 100E (on the left) and 75E25I (on the right) neuronal cultures at DIV70. A black dot and a dense band represent a detected spike and a network burst, respectively.}
\label{fig:network_images}
\end{figure*}

Human induced pluripotent stem cells (hiPSCs) \textcolor{black}{have been} differentiated in excitatory and inhibitory neurons as described in~\cite{Parodi2023}. Two different excitatory/inhibitory (E:I) configurations \textcolor{black}{have been} considered in this work:
$100\%$ excitatory (100E) and $75\%$ excitatory / $25\%$ inhibitory (75E:25I). Fig.~\ref{fig:network_images} (A-C) shows representative networks imaging for these two configurations.
Electrophysiological
 recordings \textcolor{black}{have been} conducted at 70 days in vitro (DIV) by using micro-electrode arrays (MEAs) with 60 TiN electrodes (30 $\mu$m diameter, 200 $\mu$m spacing, 8 × 8 grid), connected to the MEA2100 system (Multi Channel Systems, Reutlingen, Germany). After a 10-minute acclimation outside the incubator, spontaneous neuronal activity \textcolor{black}{has been} recorded for 15 minutes at $37$ °C and sampled at $10$ kHz. Data \textcolor{black}{have been} pre-processed by means of the peak detection algorithm devised in~\cite{Maccione2009}. Practically, by setting a differential threshold for each electrode, (8 times the standard deviation of the signal’s biological and thermal noise), the peak lifetime period (1 ms), and the refractory period (0 ms), spikes \textcolor{black}{have been} identified and used as peak trains for all the presented analyses.

\subsection{Experimental analysis}\label{sec:Exp_Analysis}

The dataset used for the analysis presented in this work consists of $n_{100E}=6$ and $n_{75E:25I}=5$ samples for purely excitatory (100E) and physiological (75E:25I) networks, respectively. After spike detection, we estimate the inter-spike
interval (ISI) – defined as the time between consecutive spikes in the network. To exclude
 coincidental simultaneous events, only ISIs greater than 1 ms \textcolor{black}{have been} considered. Given the electrode
 spacing, this threshold ensured that only spikes with a plausible causal relationship \textcolor{black}{have been} included
 in the analysis. The time-bin $t_b$ \textcolor{black}{has been} defined as the average ISI. \textcolor{black}{Tab.~\ref{tab:bin_individual} reports the values of $t_b$ for 100E and 75E:25I networks.}  

\begin{table}[h!]
\centering
\begin{tabular}{|c|c|c|} 
 \cline{2-3}
\multicolumn{1}{l|}{}& 100E & 75E:25I \\
 \hline
Sample & $t_b$ (ms) &  $t_b$ (ms)\\[0.5ex]
 \hline
$1$ & $8$ &  $12$\\
$2$ & $11$ & $6$ \\
$3$ & $10$ & $11$\\
$4$ & $9$ & $8$ \\
$5$ & $5$ &  $9$  \\
$6$ & $4$ & -- \\
 \hline
\end{tabular}
\caption{Time-bin $t_b$ of individual samples for the experiments on purely excitatory (100E) and  physiological (75E:25I) configurations.}
\label{tab:bin_individual}
\end{table}

\subsection{Avalanche detection}\label{sec:ava_detection}

To identify avalanches, neuronal activity from each MEA \textcolor{black}{has been} organized into raster plots (see Figs.~\ref{fig:network_images} D), where rows correspond to recording channels and columns to recording times. To construct the binned raster plot, the total recording time \textcolor{black}{has been} divided into contiguous, non-overlapping time windows (bins) of fixed duration $t_b$, whose values are reported in (Tab.~\ref{tab:bin_individual}), ensuring that the temporal resolution matches the intrinsic timescale of the network. For each channel and bin, the number of detected spikes \textcolor{black}{has been} counted. Bins containing at least one spike \textcolor{black}{have been} marked as active, while those with no spikes \textcolor{black}{have been} marked as inactive. This binarized representation captures the temporal structure of network activity. The overall activity at each time bin \textcolor{black}{has then been} obtained as the sum of active channels, and avalanches \textcolor{black}{have been} defined as sequences of consecutive active bins, bounded before and after by at least one inactive bin. Each avalanche is characterized by its size $S$ representing the total number of active electrodes, and its duration $T$ corresponding to the number of
 consecutive active bins multiplied by the bin size. Finally, the inter-event time $\Delta t$ \textcolor{black}{has been} defined as the
 time interval between the end of one avalanche and the start of the next one. 
\subsection{Avalanche Statistics}
The distributions of sizes $S$ and durations $T$ follow power-law distributions\cite{Beggs2003,Pasquale2008,Beggs2012}
\begin{align}
& P(S)\sim S^{-\tau}\,, \qquad P(T)\sim T^{-\alpha}\,.
\end{align}
Within the scale-free regime we also \textcolor{black}{have considered} the average size of avalanches with fixed 
duration $T$ which also follows a power-law dependence on $T$~\cite{dahmen1996hysteresis,perkovic1995avalanches,Kuntz2000}
\begin{align}
& \langle S \rangle_{S|T} \sim T^{\gamma}\,.
\end{align}
These power-law distributions appear frequently during analysis of experimental data and might not be related to the presence of a critical point~\cite{Touboul2010,Touboul2017,Priesemann2018}. Notwithstanding, as demonstrated by Sethna in 
the context of crackling noise~\cite{dahmen1996hysteresis,perkovic1995avalanches,Kuntz2000}, for critical systems the exponent $\gamma$ is related to $\alpha$ and $\tau$ by the scaling relation
\begin{align}
& \gamma=\frac{\alpha-1}{\tau-1} \,.
\label{eq:hyperscaling_relation}
\end{align}
The above relation \textcolor{black}{was} proven to hold for a wide class of experimental systems or computational models, independently of the values assumed by the two exponents $\tau$ and $\alpha$~\cite{Fontenele2019}. 

\subsection{Inter-time distribution}
The inter-time $\Delta t$ is the duration of the quiescent period that separates the end of a neuronal avalanche and the beginning of the subsequent one. The probability distribution of the inter-times $P(\Delta t)$ differs from the Poisson distribution when avalanches are correlated in time~\cite{PhysRevLett.108.228703}. In this situation, $P(\Delta t)$ might show power-law regimes $P(\Delta t)\sim \Delta t^{-\mu}$ being $\mu$ the relative exponent~\cite{lombardi2023beyond}.

\subsection{Conditional Probability Analysis}

The organization of avalanches of different size in time is evidenced by the study of the conditional probability $P(s_0,  t_0)$ of a size difference $\Delta S = S^{(i+1)}-S^{(i)}\le s_0$  given the value of the inter-time $\Delta t\le t_0$. 
Specifically, 
$P(s_0,t_0)$ is defined as 
\begin{equation}
P(s_0,t_0)=\frac{N(\Delta s<s_0;\Delta t<t_0)}{N(\Delta t<t_0)}
\label{eq:conditional_probability}
\end{equation}
where $N(\Delta t<t_0)$ is the number of event couples with temporal distance $\Delta t$ less than $t_0$ while $N(\Delta s<s_0;\Delta t<t_0)$
is the number of such couples with difference $\Delta s$ smaller than $s_0$. To quantify the strength of correlations, $P(s_0,t_0)$ \textcolor{black}{is compared} with a surrogate probability $\overline{P}(s_0,t_0)$ defined as in Eq.~\eqref{eq:conditional_probability} but for a different sequence of $\{\overline{\Delta S},\Delta t\}$. The surrogate sequence $\{\overline{\Delta S},\Delta t\}$ is obtained by reshuffling the original avalanche sizes $S$ while preserving the exact occurrence times. This ensures that the marginal distributions of sizes and inter-times coincides with the original ones, but all correlations are destroyed. 
For a large number of reshuffling $\bar{P}$ is a Gaussian with standard deviation $\sigma$. 
The difference 
\begin{equation}
    \delta P = P - \bar{P}
\end{equation}
quantifies the deviation of the real avalanche sequence from random occurrence. 
If $\delta P\le2\sigma$, the real avalanche sequence can be considered as a possible realization of random sequences and, therefore, avalanche sizes and occurrence times are uncorrelated.
For fixed $t_0$, the behavior of $\delta P$ as a function of $s_0$ provides useful information about amplification or attenuation mechanism in avalanche occurrence. A maximum 
of $\delta P$ for $\Delta S<0$ and $\Delta S>0$, indicates that, on average, avalanches close in time are attenuated or amplified in size, respectively~\cite{Lombardi2016,de2016statistical}.

\subsection{Model}\label{sec:model}
\subsubsection{Purely excitatory active Kuramoto model}
The active Kuramoto model extends the classic 
one by incorporating conservative forces into the dynamics of coupled oscillators~\cite{winfree1980geometry,shinomoto1986cooperative,shinomoto1986phase,sakaguchi1986soluble,shinomoto1986phase,kuramoto1991collective,strogatz2000kuramoto}. In the standard 
paradigm, neurons (oscillators) are characterized by their natural frequencies and coupled to all others through phase interactions. A neuron is assumed to fire whenever its phase crosses a fixed threshold (e.g. $\frac{\pi}{2}$). The active model modifies this picture by introducing a conservative force that confines the dynamics of each oscillator close to a stable-fixed point representing the neuron resting potential.
In this setting, neurons fire either when random fluctuations push them out of the stability region, or when they are triggered by interactions with connected firing neurons~\cite{buendia2021hybrid,Buenda2022}.
Formally, each neuron $j$ is represented by a variable $\theta_j$ that evolves according to~\cite{shinomoto1986cooperative,shinomoto1986phase,sakaguchi1986soluble,buendia2021hybrid}
\begin{equation}
    \dot{\theta}_{j}=
\omega_j+\xi_j-\frac{K}{N}\sum_l \sin\left(\theta_j-\theta_l\right)+a\sin(\theta_j)
\label{eq:excitatory_model}
\end{equation}
where $N$ is the number of coupled oscillators,  $\omega_j$ is an intrinsic random frequency with distribution $g(\omega)$, $K$ is the coupling strength, $a$ is the amplitude of the conservative force, and $\xi_j$ a Gaussian white noise with variance $\sigma^2_{\xi}$. The original Kuramoto model corresponds to Eq.~\eqref{eq:excitatory_model} with $a=0$. 
For $g(\omega)=\delta(\omega-\bar{\omega})$, with $\bar{\omega}$ a fix value of $\omega$, 
the system can settle close to a critical point where the oscillators are synchronized and avalanches follow power-law distributions~\cite{buendia2021hybrid,Buenda2022}. The critical point corresponds to a particular synchronization transition~\cite{buendia2021hybrid,Buenda2022}. Here, we consider $g(\omega)=\mathcal{N}(\bar{\omega},\sigma_\omega)$, i.e. a Gaussian distribution with standard deviation $\sigma_{\omega}=0.1$ and mean $\bar{\omega}$. 
For the parameters of the model, unless otherwise specified, we set $K=1$, $\bar{\omega}=1$, $a=1.055$, $\sigma_\xi=0.42$, and $N=500$. 

To perform the analysis of numerical avalanche dynamics, 
 we must first define how spikes are extracted from oscillator phases and how the corresponding raster plot is built. 
 Following~\cite{buendia2021hybrid}, we associated to each neuron with phase $\theta_j$ a function
\begin{equation}
    y_j=1+\sin(\theta_j)
\end{equation}
which spans from $0$ to $2$ for $\theta_j\in(-\pi,\pi]$.
Spikes of $j-th$ neuron are identified 
whenever $y_j$ exceeds the threshold $y_{th}=1.6$. This choice ensures that only marked variations of the oscillator trajectory are counted as firing events. 
Consistently with the analysis of experimental recordings (see Sec.~\ref{sec:Exp_Analysis}), we set the time bin $t_b$ equal to the inter spike interval. Note that in the model $t_b$ is a non-dimensional constant, its mapping to a physical time value must therefore be calibrated against experimental data (Sec.~\ref{sec:Inter-times}). 
\subsubsection{Active Kuramoto model with Inhibition Dynamics}
To take into account a second population of oscillators representing inhibitory neurons, we have extended the dynamics of Eq.~\eqref{eq:excitatory_model}. In the purely excitatory model, the interaction term tends to align the phase $\theta_{j}$ of each neuron to the average phase of the population $\Psi$, defined through the following relation
\begin{equation}
    Re^{i\Psi}=\frac{1}{N}\sum_{j=1}^{N}e^{i\theta_{j}}\,,
\end{equation}
where $R$ is the norm of the Kuramoto-Daido order parameters~\cite{strogatz2000kuramoto}.
Accordingly, Eq.~\eqref{eq:excitatory_model} can be rewritten in terms of $R$ and $\Psi$ as
\begin{equation}
    \dot{\theta}_{j}=
\omega_j+\xi_j-KR\sin\left(\theta_j-\Psi\right)+a\sin(\theta_j)\,\,
\label{eq:mean_field_excitatory_model}
\end{equation}
which makes the mean-field alignment nature of the interaction explicit~\cite{kuramoto1991collective,strogatz2000kuramoto,shinomoto1986cooperative,acebron2005kuramoto}. 
Many theoretical and numerical studies have generalized Eq.~\eqref{eq:mean_field_excitatory_model} to multi-population systems~\cite{sakaguchi1986soluble,strogatz2000kuramoto,shinomoto1986cooperative,acebron2005kuramoto,Montbri2018,winfree1980geometry}.
In such models, the interactions between different populations have the same form as in Eq.~\eqref{eq:mean_field_excitatory_model}, but with distinct coupling constants $K$, which may also take negative values. While this assumption is reasonable for excitatory neurons, it is not biologically appropriate for an inhibitory population. 
Indeed, when an inhibitory neuron fires, the total activity of the network must decrease. Conversely, it has been shown that this type of inhibition interaction might enhance activity and synchronization~\cite{Shimokawa2006,Montbri2018}. 
This unrealistic scenario occurs because inhibition is modeled as a purely repulsive force, which may induce excitatory neurons to spike when the inhibitory oscillators' phase is close to the stable position of the excitatory ones.

To heal the contradiction with the behavior of real neurons, we have modified the conservative potential so that its amplitude depends on the activity of the inhibitory population. Defining the phases of a generic excitatory or inhibitory neuron as $\theta^{(E)}_{j}$ and $\theta^{(I)}_{j}$, their evolution equations are:
\begin{widetext}
\begin{align}
    &\dot{\theta}_{j}^{E}=
\omega_j+\xi_j-K_ER_E\sin\left(\theta_j^E-\Psi_E\right)+\left[a+K_IR_I^2\cos^2(\Psi_I)\right]\sin(\theta_j^E)\nonumber\\
&\dot{\theta}_{j}^{I}=
\omega_j+\xi_j-K_ER_E\sin\left(\theta_j^I-\Psi_E\right)+\left[a+K_IR_I^2\cos^2(\Psi_I)\right]\sin(\theta_j^I)
\label{eq:inhibitory_model}
\end{align}
\end{widetext}
where $R_E$, $R_I$, $\Psi_E$ and $\Psi_I$ are defined by
\begin{align}
    &R_Ee^{i\Psi_E}=\frac{1}{N_E}\sum_{j=1}^{N_E}e^{i\theta_{j}^{E}}\,,\nonumber\\
    &R_Ie^{i\Psi_I}=\frac{1}{N_I}\sum_{j=1}^{N_I}e^{i\theta_{j}^{I}}\,.
\end{align}
In Eq.~\eqref{eq:inhibitory_model}, each neuron tends to align its phase with the average activity of the excitatory population, while inhibitory activity modulates the conservative potential. This is implemented by the terms in square brackets, which effectively drive all neurons towards their quiescent state depending on the average activity of inhibitory neurons, reducing their firing probability. 

By setting $K_E =K\frac{N_E}{N}$ and $K_I = K\frac{N_I}{N}$, the model reduces to Eq.~\eqref{eq:mean_field_excitatory_model} when $N_E = N$. 
From the biological point of view, the inhibition dynamics implemented here captures the regulatory effect of inhibition in balancing excitation. In different models~\cite{diSanto2018,PhysRevResearch.2.013318}, this stabilization is introduced by synaptic resource depletion only, but here we focus exclusively on inhibition dynamics and do not include explicitly the synaptic fatigue. 
The phase diagram of this model as a function of the amplitude of noise $\sigma_\xi$ and the amplitude of the conservative potential $a$ (not shown here) closely resembles the one already obtained for the purely excitatory case in~\cite{buendia2021hybrid}. The parameters have been set as follows: $K=1$, $\bar{\omega}=1$,$\sigma_\omega=0.05$, $a=0.92$, $\sigma_\xi=0.35$, and $N=750$. 
To determine the avalanche dynamics in this case, we apply the same procedure described above for the single-population model. 

\subsubsection{External Poisson noise}
The variability among different experimental samples within a given network class (i.e., 100E or 75E:25I) not only leads to different scaling exponents but can also be quantified by examining the Fano Factor, that is the ratio between the variance and the mean of the number of spikes within a given time window. For the purely excitatory networks, there is a positive correlation between the exponents and this measure of variability (see SM~\cite{supp}). Conversely, for the 75E:25I networks, the measures of this ratio are too noisy to reliably assess the variability.
To reproduce this variability in numerical simulations, we have included an additional noise source into the model. Noise is treated as an external contribution superposed to the simulated raster plot. 
In particular, in each channel and to each bin we have added an entry that has probability $p$ of being active and probability $1-p$ of being inactive. 
The external noise has a double effect. On one side, it reduces the probability of small avalanches by merging avalanches together. On the other hand, it reduces the inter-spike time, and therefore the time-bin $t_b$, used for the avalanche detection, fragmenting large avalanches into smaller ones. 
In both single- and two-population models, two different noise levels have been used ($p=10^{-5}$ and $p= 10^{-4}$ for 100E, $p=10^{-4}$ and $p=5\cdot 10^{-4}$ for 75E:25I).

\subsection{Power-law fitting and error estimation}
Power-law exponents for avalanche size, duration, and inter-event interval distributions \textcolor{black}{have been} estimated using least-squares fitting in log–log scale. Empirical distributions \textcolor{black}{have been} first log-binned to reduce fluctuations in the tails. Linear regression \textcolor{black}{has been} then applied on the log-binned probability density function. Fitting \textcolor{black}{has been} restricted to the scaling regime, namely selected linear regions of the
log–log plots, excluding the peak associated with synchronized bursting and the noisy tail at large event sizes. 
To assess the robustness of the fitted exponents, we \textcolor{black}{have systematically varied} the fitting window. The reported uncertainty \textcolor{black}{has been} taken as 
half the maximum variation observed across the tested ranges. 
This approach reflects how sensitive the fitted exponent is to the selected bounds and provides a conservative estimate of
the exponents in data with non-ideal power-law. 
Maximum likelihood methods and goodness-of-fit tests \textcolor{black}{have not been} applied, because bimodality violates their assumptions. The error on derived quantities, as the exponent $\gamma$ estimated by Eq.~\eqref{eq:hyperscaling_relation}, \textcolor{black}{has been} computed using the standard error propagation formula. 
The error on crossover time estimated from the distribution of inter-times $P(\Delta t)$ \textcolor{black}{has been} computed from the intersection between the two power-laws by varying the fitting window. 
The difference of conditional probabilities $\delta P$, instead, has been computed as an average over $M=10^3$ surrogate distributions while the error bars have been estimated as $3$ standard deviations of the Gaussian distribution obtained for reshuffled data. 
Finally, the crossover time for $\delta P$ \textcolor{black}{has been} estimated from the plots. An uncertainty of $5t_b$ \textcolor{black}{has been} assigned, reflecting the variability in identifying the crossover curve.

\section{Results}\label{sec:Results}

\subsection{Avalanches}\label{sec:Avalanches}

\begin{figure*}[ht!]
    \centering
    \includegraphics[width=1.0\textwidth]{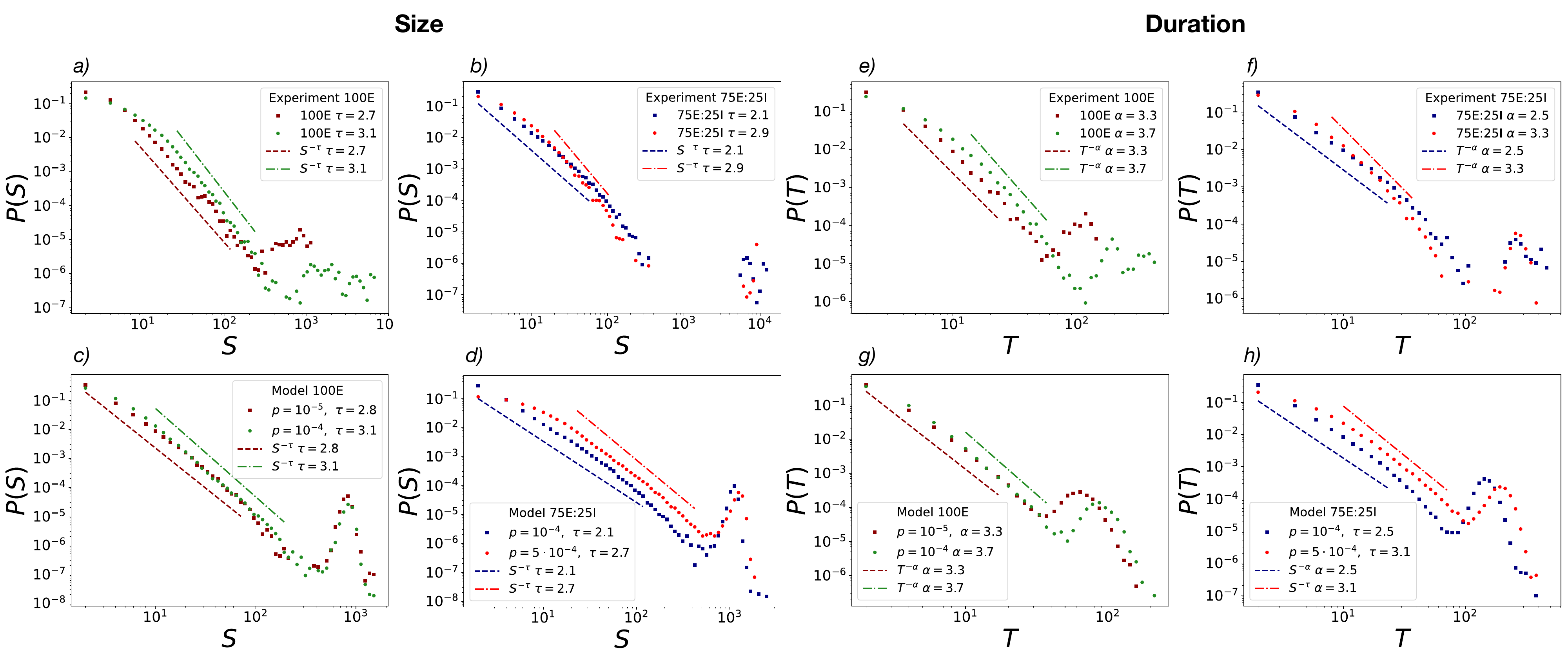}

    \caption{Size and duration distributions for experimental and numerical data with superimposed theoretical power-law behavior (dashed and dot-dashed lines). Circles and squares represent samples with high or low values of the exponents $\tau$ and $\alpha$, respectively. a) Size distribution of experiments on purely excitatory networks. b) Size distribution of experiments on physiological networks. 
    c-d) Size distributions of single- and two- population active Kuramoto model with inhibition dynamics. 
    e) Duration distribution of experiments on purely excitatory networks. f) Duration distribution of experiments on physiological networks. 
    g-h) Duration distributions of single- and two- population active Kuramoto model with inhibition dynamics.}
\label{fig:size_duration_distributions}
\end{figure*}

We investigate the properties of neuronal avalanches in electrophysiological recordings of hiPSC-
derived cortical cultures with different excitatory/inhibitory (E/I) ratios: purely excitatory (100E) and physiological (75E:25I) networks. The avalanche size and duration distributions (Fig.~\ref{fig:size_duration_distributions} (a), (b), (e), (f)), exhibit
bimodal shapes. In particular, large and long-lasting avalanches associated with network-level
 synchronization phenomena show narrow distributions, while avalanches of intermediate size follow
a power-law pattern extending for nearly two decades.
Analysis of individual samples (Tab.~\ref{tab:results_individual_100E}) reveals that both 100E and 75E:25I networks can be divided into two subgroups based on their scaling exponent values for avalanche size
and duration distributions.
Average group distributions indicate the existence of an intermediate scaling regime with exponents $\tau$ and $\alpha$ which are larger for higher values of the Fano factor (see SM~\cite{supp}). 

Interestingly, the decrease of the distributions is steeper for fully excitatory networks compared to physiological ones. 
These results can be explained in terms of the experimental observation~\cite{Parodi2023}, evidencing that in fully excitatory networks synchronization occurs with a higher frequency with respect to physiological cultures. Therefore, synchronization hampers the occurrence of large avalanches in favour of bursts encompassing the entire system. 

\begin{table}[h!]
\centering
\begin{tabular}{|c|c|c|c|c|} 
 \cline{2-5}
\multicolumn{1}{l|}{}& \multicolumn{2}{c|}{ 100E} & \multicolumn{2}{c|}{ 75E:25I} \\
\hline
Sample & $\tau$ & $\alpha$ & $\tau$ & $\alpha$ \\[0.5ex]
 \hline
$1$ &  $3.3\pm 0.2$ & $3.6\pm0.3$ & $2.9\pm 0.2$ & $3.4\pm0.2$ \\
$2$ &  $2.8\pm 0.1$ & $3.2\pm0.2$ & $3.0\pm 0.2$ & $3.3\pm0.3$\\
$3$ &  $2.8\pm 0.1$ & $3.3\pm0.2$ & $2.1\pm0.1$ & $2.5\pm0.2$\\
$4$ &  $3.3\pm0.2$ & $3.7\pm0.3$ & $2.0\pm0.1$ & $2.6\pm0.2$\\
$5$ &  $3.2\pm0.2$ & $3.6\pm0.3$ & $2.0\pm0.2$ & $2.4\pm0.2$\\
$6$ &  $3.2\pm0.2$ & $3.5\pm0.3$ & -- & --\\
 \hline
\end{tabular}
\caption{Critical exponents of the avalanche distributions for individual samples for the experiments on purely excitatory (100E) and physiological (75E:25I) networks.}
\label{tab:results_individual_100E}
\end{table}

Notably, the exponents $\tau$ and $\alpha$ 
reported in Tab.~\ref{tab:average_exponent}, are larger than those usually found in literature for spontaneous activity in neuronal cultures and not compatible with those of the universality class of the mean-field branching process ($\tau=\frac{3}{2}$ and $\alpha=2$)~\cite{Beggs2003,Beggs2012,deArcangelis2014}. 
An emerging paradigm in brain dynamics is the concept of self-organized bistability~\cite{PhysRevResearch.2.013318}, which suggests that neural networks operate near a synchronization transition, where scale-free avalanches can arise~\cite{diSanto2018,PhysRevResearch.2.013318,buendia2021hybrid}. To explore this, we employ the active rotator model generalizing the Kuramoto oscillators introducing an excitable behavior~\cite{sakaguchi1986soluble,sakaguchi1988phase}. This excitability drives the system activity to organize into avalanches whose size and duration distributions exhibit bimodality, consisting of a power-law regime followed by a bump associated with synchronized bursting~\cite{PhysRevResearch.2.013318,buendia2021hybrid}. We extend the active Kuramoto model in two ways (see Sec.~\ref{sec:model}). First, we incorporate an external noise to reproduce the experimental variability of exponents. Second, we develop a two-population version of the model to account explicitly for the effects of the inhibitory population through a biological grounded modulation of the conservative potential.
As shown in
Fig.~\ref{fig:size_duration_distributions}, both the single- and two-population models correctly reproduce the bimodal distributions observed experimentally: an initial power law regime followed by a bump at large $S$ and $T$ corresponding to synchronization phenomena. Moreover, tuning the external noise allows reproducing the variability in exponent values (Tab.~\ref{tab:average_exponent}). 

\begin{figure*}[ht!]
    \centering
    \includegraphics[width=1.0\textwidth]{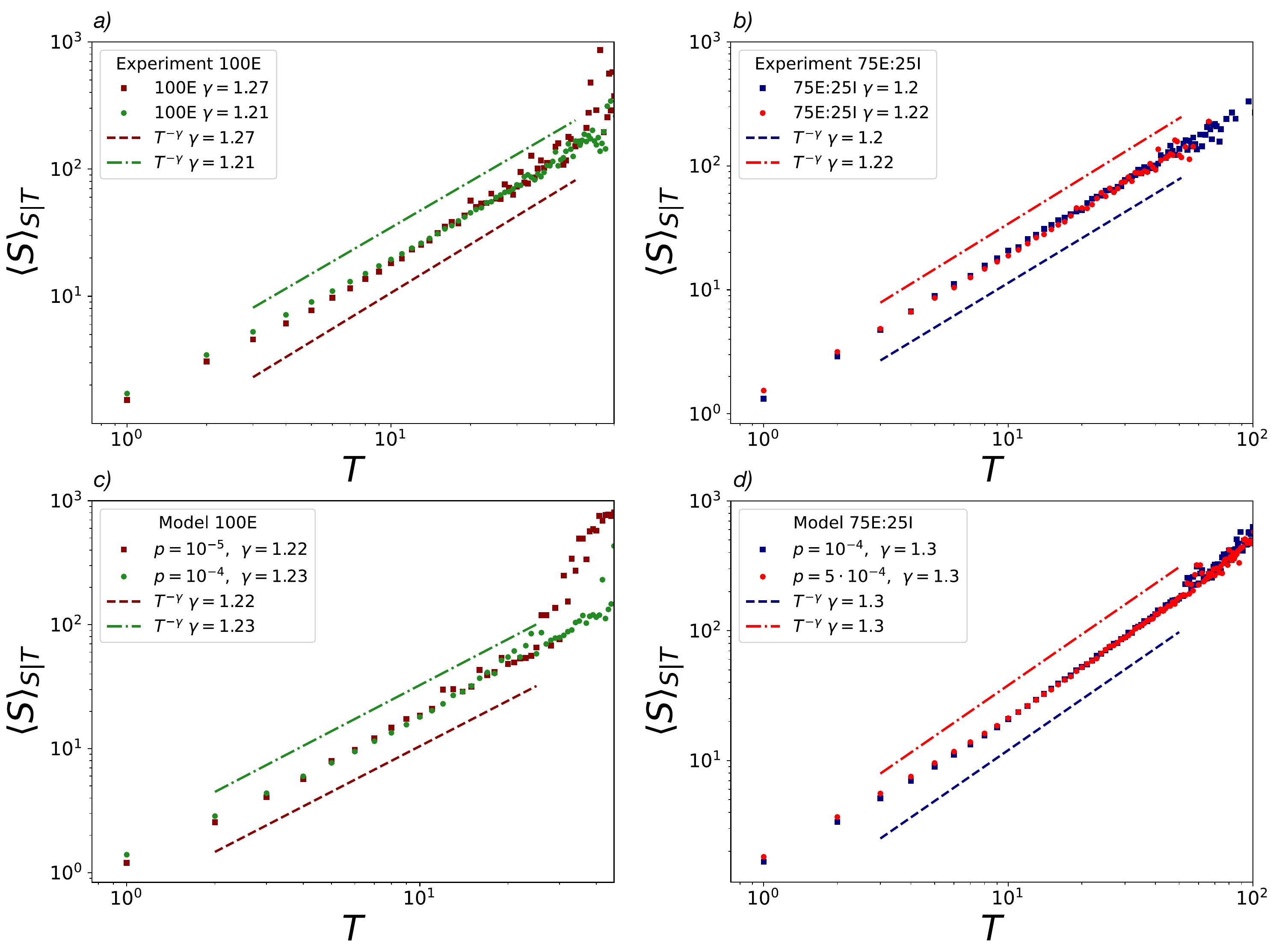}

    \caption{Conditional expectation of avalanche size $\langle S \rangle_{S|T}$ for experimental and numerical data with superimposed theoretical power-law behavior (dashed and dot-dashed lines). Circles and squares represent groups with high or low values of the exponents $\tau$ and $\alpha$, respectively. a) Experiments on purely excitatory networks. b) Experiments on physiological networks. 
    c-d) Single- and two- population active Kuramoto model with inhibition dynamics.}
\label{fig:average_size}
\end{figure*}

To further investigate the power-law activity regimes, we also analyze the behavior of the average size of avalanches conditioned on the duration $T$, $\langle S \rangle_{S|T}$, and compare the results of experimental and simulated data (Fig.~\ref{fig:average_size}). 
The relation between $\langle S \rangle_{S|T}$ and $T$ deviates from the power law behavior when the bursts at large $S$ and $T$ are approached, in agreement with the Sethna theory holding only at the critical point, here in the scaling regime. 

\begin{table}[h!]
\centering
\begin{tabular}{ |c|c|c|c|c| } 
 \hline
Data & $\tau$ & $\alpha$ & $\gamma$ & $\gamma$ (Eq.~\eqref{eq:hyperscaling_relation}) \\[0.5ex]
 \hline
Exp $100$E & $2.7\pm 0.2$ & $3.3\pm0.2$ & $1.3\pm0.1$ & $1.3\pm0.2$\\
$p=10^{-5}$ $100$E & $2.8\pm0.2$ & $3.3\pm0.2$ & $1.2\pm0.1$ & $1.3\pm0.2$ \\
\hline
Exp $100$E & $3.1\pm 0.2$ & $3.7\pm0.3$ & $1.2\pm0.1$ & $1.3\pm0.2$ \\
$p=10^{-4}$ $100$E & $3.1\pm0.3$ & $3.7\pm0.2$ & $1.2\pm0.1$ & $1.3\pm0.2$ \\
\hline
Exp $75$E:$25$I & $2.1\pm0.2$ & $2.5\pm0.2$ & $1.2\pm0.1$ & $1.3\pm0.3$ \\
$p=10^{-4}$ $75$E:$25$I & $2.1\pm0.3$ & $2.5\pm0.2$ & $1.3\pm0.2$ & $1.3\pm0.3$ \\
\hline
Exp $75$E:$25$I & $2.9\pm0.2$ & $3.3\pm0.2$ & $1.2\pm0.1$ & $1.2\pm0.2$ \\
$p=5\cdot10^{-4}$ $75$E:$25$I & $2.7\pm0.3$ & $3.1\pm0.2$ & $1.3\pm0.2$ & $1.2\pm0.3$ \\
 \hline
\end{tabular}
\caption{Exponents of size ($\tau$) and duration ($\alpha$) distributions and of conditional expectation $\langle S \rangle_{S|T}$ ($\gamma$) for different data-sets. Except for the last column, where the scaling relation~\eqref{eq:hyperscaling_relation} has been used, the values refer to results obtained from a fitting procedure (see Sec.~\ref{sec:Material_Method}).}
\label{tab:average_exponent}
\end{table}

Moreover, for both experimental and numerical networks, the exponent $\gamma$ shows minimal variations between purely excitatory and physiological networks (Figs.~\ref{fig:average_size}). 
These considerations further support our hypothesis that differences between groups within the same experimental setup are akin to some form of noise.
As explained above, the effect of the noise mostly results in larger exponents.
Interestingly, the values of $\gamma$ determined by the fitting procedure are coherent with those estimated from the scaling relation~\eqref{eq:hyperscaling_relation}. Moreover, those values are compatible with those reported in~\cite{Fontenele2019}. 

\begin{figure*}[ht!]
    \centering
    \includegraphics[width=1.0\textwidth]{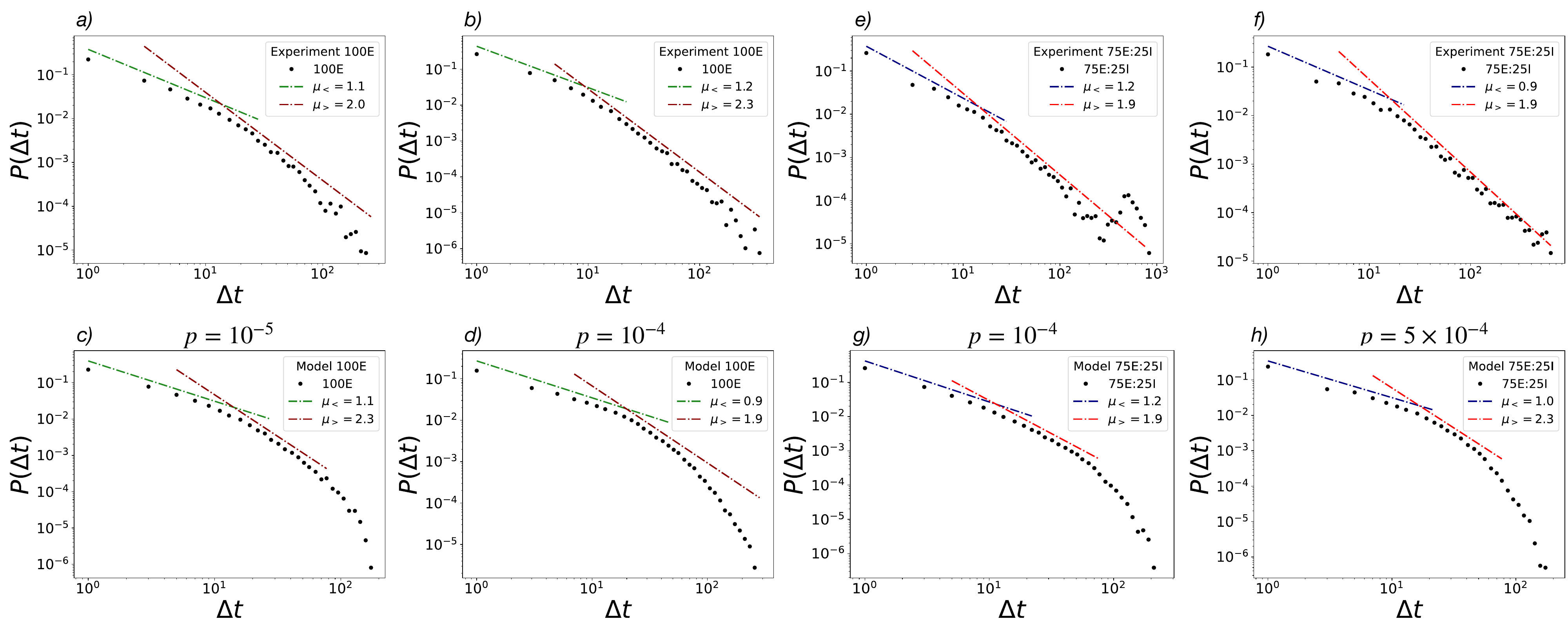}
    \caption{Inter-time distributions with superimposed theoretical power-law behavior. The crossover time can be estimated from the intersection between the two power-law regimes. a-b) Experiments on purely excitatory networks. e-f) Experiments on physiological networks. 
    c-d-g-h) Single- and two- population active Kuramoto model with inhibition dynamics.}
\label{fig:inter-time_distribution}
\end{figure*}

\subsection{Inter-times}\label{sec:Inter-times}

To probe the temporal correlations of activity, we measure the distribution of inter-times $\Delta t$ between successive avalanches. This distribution is able to discriminate between a Poisson process, in which case it exhibits an exponential behavior, and a process with temporal correlations if a power-law behavior is detected. Assuming that small-size avalanches are mostly dominated by external noise, we restrict our analysis to sizes exceeding a given threshold $S^{\star}$, as in~\cite{bianchi2018critical}. We consider two values of the threshold per network type, corresponding to the two subgroups identified earlier. The experimental inter-event interval distributions, in units of $t_b$ and averaged over the samples of each subgroup, exhibit a clear crossover between two power-law regimes (Fig.~\ref{fig:inter-time_distribution} (a-b-e-f)). Importantly, the corresponding distributions $P(\Delta t)$ for both single- and two-population models closely match the experimental findings, including the crossover between the two power-law regimes (Fig.~\ref{fig:inter-time_distribution} (c-d-g-h)). Therefore, for both experimental and numerical data, the distributions indicate the presence of complex temporal correlations resulting in the crossover (Fig.~\ref{fig:inter-time_distribution}). 
The two exponents $\mu_<$ (for small $\Delta t$) and $\mu_>$ (for large $\Delta t$) are consistent from sample to sample (Tabs.~\ref{tab:inter-time_individual_75-25}-\ref{tab:exponent_inter-time_crossover}), since size thresholding reduces the effect of external noise. 

\begin{table}[h!]
\centering
\begin{tabular}{|c|c|c|c|c|c|c|}
 \cline{2-7}
\multicolumn{1}{l|}{}& \multicolumn{3}{c|}{ 100E} & \multicolumn{3}{c|}{ 75E:25I} \\
\hline
Sample & $\mu_<$ & $\mu_>$ & $S^\star$ & $\mu_<$ & $\mu_>$ & $S^\star$\\[0.5ex]
 \hline
$1$ & $1.1\pm0.3$ & $1.9\pm0.2$ & $5$ & $0.9\pm0.2$ & $2.0\pm0.3$ & $9$\\
$2$ & $1.1\pm0.2$ & $2.0\pm0.3$ & $4$ & $0.9\pm0.2$ & $1.9\pm0.2$ & $9$\\
$3$ & $1.0\pm0.2$ & $2.0\pm0.3$ & $4$ & $1.2\pm0.2$ & $2.0\pm0.3$ & $8$\\
$4$ & $1.1\pm0.3$ & $2.3\pm0.3$ &  $5$ & $1.1\pm0.2$ & $1.9\pm0.2$  & $8$\\
$5$ & $1.2\pm0.3$ & $2.3\pm0.4$ &  $5$  & $1.1\pm0.2$ & $2.0\pm0.2$ & $8$ \\
$6$ & $1.1\pm0.2$ & $2.4\pm0.4$ & $5$ & -- & -- & -- \\
 \hline
\end{tabular}
\caption{Inter-time distribution exponents $\mu_<$ and $\mu_>$ resulting from the analysis of individual samples for experiments on purely excitatory (100E) and physiological (75E:25I) networks.}
\label{tab:inter-time_individual_75-25}
\end{table}

\begin{table}[h!]
\centering
\begin{tabular}{ |c|c|c| } 
 \hline
Data & $\mu_<$ & $\mu_>$ \\[0.5ex]
 \hline
Exp $100$E & $1.1\pm 0.2$ & $2.0\pm0.3$ \\
$p=10^{-5}$ $100$E & $1.1\pm0.2$ & $2.3\pm0.3$ \\
\hline
Exp $100$E & $1.2\pm 0.2$ & $2.3\pm0.2$ \\
$p=10^{-4}$ $100$E & $0.9\pm0.2$ & $1.9\pm0.3$ \\
\hline
Exp $75$E$25$I & $1.2\pm0.1$ & $1.9\pm0.2$ \\
$p=10^{-4}$ $75$E$25$I & $1.2\pm0.2$ & $1.9\pm0.3$ \\
\hline
Exp $75$E$25$I & $0.9\pm0.1$ & $1.9\pm0.2$ \\
$p=5\cdot10^{-4}$ $75$E$25$I & $1.0\pm0.2$ & $2.3\pm0.3$\\
\hline
\end{tabular}
\caption{Exponents $\mu_<$ and $\mu_>$ of the average inter-time distribution.} 
\label{tab:exponent_inter-time_crossover}
\end{table}

To better quantify temporal correlations, we estimate the crossover time $\mathcal{T}_I$ in ms for each sample (Tab.~\ref{tab:crossover_individual_100}).  Purely excitatory networks exhibit shorter $\mathcal{T}_I$ than physiological ones, consistent with oscillations in the alpha band. 

\begin{table}[h!]
\centering
\begin{tabular}{ |c|c|c|c|c| } 
 \cline{2-5}
\multicolumn{1}{l|}{} & \multicolumn{2}{c|}{ 100E} & \multicolumn{2}{c|}{ 75E:25I} \\
 \hline
Sample & $\mathcal{T}_I$ (ms) & $\mathcal{T}_C$ (ms) & $\mathcal{T}_I$ (ms) & $\mathcal{T}_C$ (ms)\\[0.5ex]
 \hline
$1$ &  $90\pm 50$ & $80\pm40$ &  $200\pm 70$ & $180\pm60$ \\
$2$ &  $130\pm 90$ & $110\pm60$ &  $110\pm 30$ & $90\pm30$\\
$3$ &  $140\pm80$ & $100\pm50$ &  $230\pm30$ & $170\pm60$ \\
$4$ &  $100\pm50$ & $90\pm50$ & $100\pm60$ & $120\pm40$\\
$5$ &  $40\pm30$ & $40\pm20$ &  $130\pm50$ & $100\pm50$\\
$6$ &  $30\pm10$ & $30\pm20$ & -- & --\\
 \hline
\end{tabular}
\caption{Crossover time of the inter-time distributions $\mathcal{T}_I$ and of the conditional probability differences $\mathcal{T}_C$ for individual samples for the experiments on purely excitatory (100E) and physiological networks (75E:25I).}
\label{tab:crossover_individual_100}
\end{table}

\begin{table}[h!]
\centering
\begin{tabular}{ |c|c|c|c| } 
 \hline
Data &  $\mathcal{T}_I$ (ms) & $\mathcal{T}_C$ (ms)  \\[0.5ex]
 \hline
Exp $100$E & $135\pm 60$ & $105\pm40$ \\
$p=10^{-5}$ $100$E & $145\pm 80$ & $75\pm30$ \\
\hline
Exp $100$E & $65\pm 20$ & $60\pm20$ \\
$p=10^{-4}$ $100$E & $125\pm 50$ & $60\pm20$ \\
\hline
Exp $75$E$25$I & $155\pm30$ & $130\pm30$ \\
$p=10^{-4}$ $75$E$25$I & $125\pm90$ & $75\pm30$ \\
\hline
Exp $75$E$25$I & $155\pm40$ & $135\pm40$ \\
$p=5\cdot10^{-4}$ $75$E$25$I & $175\pm70$ & $70\pm30$ \\
\hline
\end{tabular}
\caption{Comparison between the average crossover times extrapolated from the inter-time distributions $\mathcal{T}_I$ and from the conditional probability differences $\mathcal{T}_C$ for experimental and numerical data. The adimensional crossover times for numerical data are $14\pm8$ and $19\pm7$ (100E) and $13\pm10$ and $19\pm 8$ (75E:25I) for the low and high noise cases, respectively.} 
\label{tab:comparison-time_bin_fit}
\end{table}

To compare these results with numerical simulations, a physical timescale must be assigned to the model. Our choice is to match the time-bin of the model to the average experimental time-bin, where the average is performed over the respective subgroups (more details in SM~\cite{supp}). Thus, it is possible to express the crossover time $\mathcal{T}_I$ in ms. The average experimental crossover-times $\mathcal{T}_I$ and the corresponding values computed from simulations are reported in Tab.~\ref{tab:comparison-time_bin_fit}. All model-derived crossover-times $\mathcal{T}_I$, except for the purely excitatory model with $p=10^{-4}$, are compatible with their experimental counterparts. The discrepancy observed for the 100E case is probably related to an overestimation of the average time-bin $t_b$ of the corresponding experiment (see SM~\cite{supp}).  



\begin{figure*}[ht!]
    \centering
    \includegraphics[width=1\textwidth]{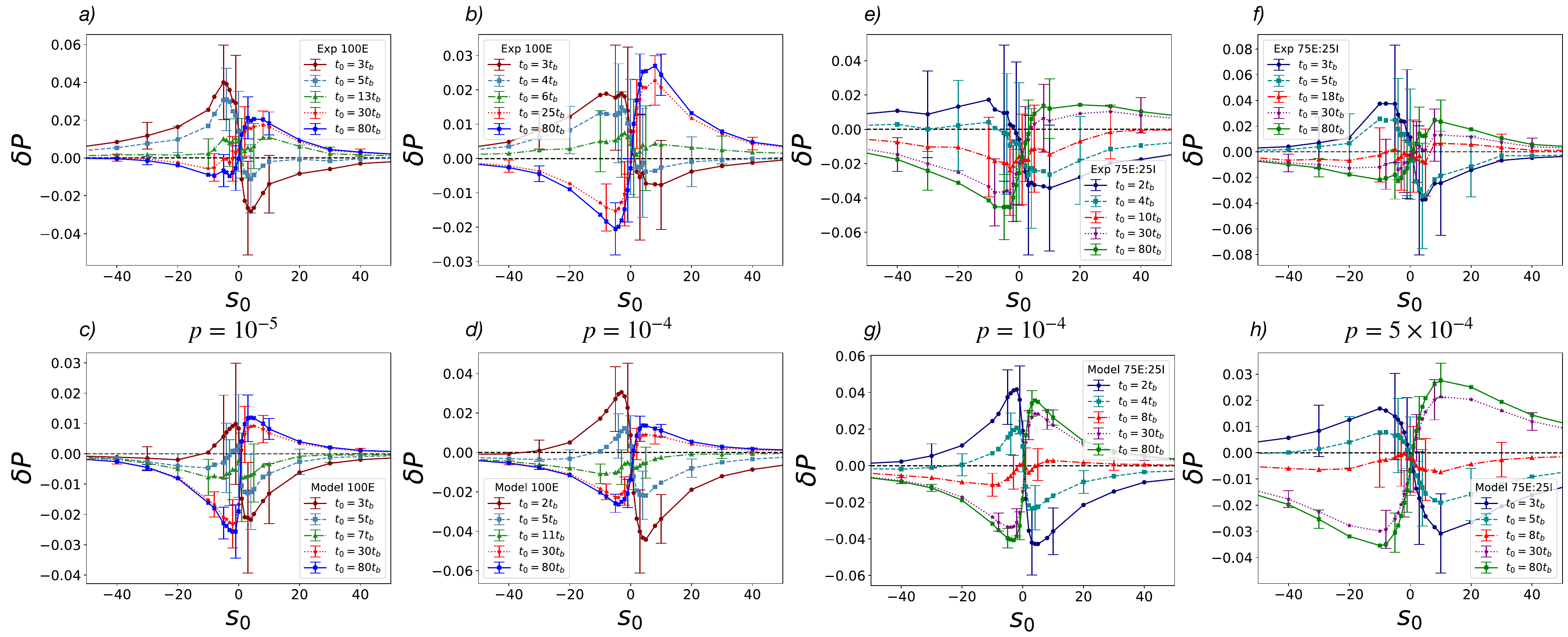}

    \caption{Difference $\delta P$ between conditional probabilities of observing a difference between subsequent avalanche sizes smaller than a given value $s_0$ conditioned on the fact that their inter-time is lower than $t_0$. The crossover time can be estimated as the time when the switching between amplification and attenuation regime occurs. a-b) Experiments on purely excitatory networks. e-f) Experiments on physiological networks. 
    c-d-g-h) Single- and two- population active Kuramoto model with inhibition dynamics.}
\label{fig:conditional_probabilities}
\end{figure*}

\subsection{Correlation between avalanche size and occurrence time}\label{sec:conditional_probabilities}
Finally, we investigate the correlations between successive avalanches by analyzing the conditional probability $P(s_0, t_0)$ to observe two consecutive  avalanches with a size difference $\Delta S = S^{(i+1)}-S^{(i)}\le s_0 $  given the value of their inter-time $\Delta t\le t_0$. More specifically, we consider the difference between conditional probabilities \begin{equation}
\delta P(s_0, t_0)=P(s_0, t_0)-\bar{P}(s_0,t_0)
\end{equation}
where $P(s_0, t_0)$ and $\bar{P}(s_0, t_0)$ are the conditional probabilities computed on the original or reshuffled  avalanche sequence, respectively (see Sec.~\ref{sec:Material_Method}). A careful analysis of this quantity can reveal correlations between differences in avalanche size and corresponding quiescent times~\cite{deArcangelis2014,Lombardi2016,de2016statistical,lombardi2023beyond}.
In all experimental samples the quantity $\delta P$ exhibits a complex scenario (see SM~\cite{supp}).  
In Fig.~\ref{fig:conditional_probabilities} (a-b-e-f), the average curves for each subgroup networks clearly display a crossover between different regimes, also reproduced by the Kuramoto models (Fig.~\ref{fig:conditional_probabilities} c-d-g-h). For small values of $t_0$, $\delta P$ presents a positive maximum for $s_0<0$ and a negative minimum for $s_0>0$, indicating that for avalanches close in time, i.e. occurring after a short quiescence period, the second avalanche tends to be smaller than the preceding one (attenuation regime).
For large values of $t_0$, $\delta P$ has the opposite behavior with its maximum at  positive values of $s_0$, implying that for avalanches distant in time, the second avalanche tends to be larger than the preceding one (amplification regime). By varying $t_0$, $\delta P$ changes continuously between the two different regimes,  becoming almost flat at a particular value of $t_0$, identifying the crossover time $\mathcal{T}_C$ between the attenuation and amplification dynamics. 

For each individual sample, we estimate the crossover time $\mathcal{T}_C$ and we report the results in Tab.~\ref{tab:crossover_individual_100}.
We notice that $\mathcal{T}_C$ is smaller for purely excitatory networks than physiological ones. Moreover,  $\mathcal{T}_I$ obtained from the inter-time distributions and $\mathcal{T}_C$ from the crossover of conditional probability differences are comparable within the error bars. Let us also stress that the same kind of observations can be made for numerical data, as can be seen in Tabs.~\ref{tab:comparison-time_bin_fit}. 

\section{Discussion}\label{sec:discussion}
Our results demonstrate that hiPSC-derived neuronal networks display the coexistence of scale free avalanches with strongly synchronized bursts, and that these features are reproduced by our novel Kuramoto model. Taken together, the experiments and theory support the view that neuronal systems operate in a regime of self-organized bistability, at the boundary of synchronization where power-law avalanche regimes and oscillatory events naturally emerge~\cite{diSanto2018,buendia2021hybrid}.
This behavior differs from what already found in in vitro models of cortical
murine networks: in the seminal works in acute cortical slices~\cite{Beggs2003} and some years later in dissociated cortical cultures~\cite{Pasquale2008}, a
power-law behavior in the avalanches’ distributions of size and lifetime was found. The presence of
a bump of activity in correspondence of big avalanches was achieved in such experimental models
by altering the excitatory/inhibitory balance of the networks with the administration of GABA
antagonists like bicuculline or picrotoxin. In the recordings analyzed in this work, where neurons
derive from hiPSCs, the bistable dynamics (although with different exponents) persists in both
physiological (75E:25I) and fully excitatory networks (100E). This evidence could be explained by
speculating a peculiar network connectivity for hiPSCs derived neuronal networks: as
computational studies proved~\cite{pajevic2009efficient,massobrio2015self} the complexity of the network organization pushes the network dynamics towards
power-law and/or bistable dynamics: Massobrio, et al. simulated a power law regime only with a
network connectivity organized with the existence of hubs and clusters of neurons. This feature
cannot be explained by the computational choice pursued in this work: the Kuramoto model
implements a full connectivity matrix, thus ignoring possible connectivity rules. 

The central innovation of this work is the explicit incorporation of inhibition dynamics into the Kuramoto framework. Previous oscillator models successfully capture synchronization phenomena\cite{montbrio2004synchronization,Montbri2018} but represent inhibition in a manner that results in a non-biological enhancement of activity~\cite{Shimokawa2006}. Here, instead, we implement the inhibition dynamics by letting the amplitude of the conservative potential be dependent on the inhibitory activity, establishing a biologically grounded mechanism of regulation that accounts for the coexistence of avalanche-dominated and synchronized states. 
We also found that an external Poisson noise reproduces the variability observed across experimental samples. This result is consistent with previous reports showing that apparent critical scaling might emerge from stochastic variability~\cite{Touboul2010,Touboul2017,Priesemann2018} and that noise actively modulates avalanche dynamics~\cite{Lombardi2016}. 

As discussed in Sec.~\ref{sec:Avalanches}, the exponents $\tau$ and $\alpha$ characterizing the avalanche size and duration distributions, measured in both experiments and simulations, differ from those predicted by the mean-field branching process. Consequently, the scaling exponent $\gamma$ also deviates from its theoretical value $\gamma=2$.
Previous studies~\cite{Neto2022} have attributed such discrepancies to subsampling effects in experimental recordings. While this explanation may account for the experimental data, it does not explain the behavior of our model. 
Moreover, the values of $\gamma$ estimated in Sec.~\ref{sec:Avalanches} are consistent with those reported in~\cite{Fontenele2019}, where the exponents $\tau$ and $\alpha$ likewise differ from the mean-field predictions.
Beyond static avalanche statistics~\cite{Beggs2003,Pasquale2008,Beggs2012,deArcangelis2014}, the analysis of inter-time distributions and conditional probabilities has unveiled temporal correlations between consecutive avalanches. The inter-time distributions exhibited a crossover between two different power-law regimes. This feature has already been observed in human brain, where the crossover is related to the $\alpha$ rhythm~\cite{lombardi2023beyond}. Interestingly, also in our study this crossover falls into the alpha-band for both purely excitatory and physiological networks. The analysis of conditional probabilities further confirms the existence of a crossover between attenuation and amplification regimes, with a characteristic time-scale that closely matches the crossover-time estimated from the inter-event time distributions, as expected~\cite{lombardi2023beyond}.
The ability of the model to reproduce not only the synchronized activity but also the avalanche dynamics, including these memory-like effects strongly suggests self-organized bistability as a unifying paradigm for brain dynamics, linking avalanche propagation, synchronization, and temporal correlations within a single theoretical construct.

\begin{acknowledgments}
 The authors acknowledge funding from NEXTGENERATIONEU (NGEU) funded by the Ministry of University and Research (MUR), National Recovery and Resilience Plan (NRRP), and Project MNESYS (PE0000006)-A multiscale integrated approach to the study of the nervous system in health and disease (DN. 1553 11.10.2022). AS acknowledges funding from the Italian Ministero dell’Università e della Ricerca under the programme PRIN 2022 (”re-ranking of the final lists”),
number 2022KWTEB7, cup B53C24006470006. 
\end{acknowledgments}
\bibliography{bibliography}

\end{document}


\title{Supplemental Material\\Novel Kuramoto model with inhibition dynamics modeling critical dynamics and synchronization in neuronal cultures}

\author{D. Lucente$^{1}$}
\altaffiliation[]{These authors have contributed equally to this study.}

\author{L. Cerutti$^{2,3}$}
\altaffiliation[]{These authors have contributed equally to this study.}

\author{M. Brofiga$^{2,3}$}  \author{A. Sarracino$^{4}$} \author{G. Parodi$^{2,3}$} \author{S. Martinoia$^{2,5,6}$} \author{P. Massobrio$^{2,7}$}\altaffiliation[]{These authors jointly supervised the study.}
\author{L. de Arcangelis$^{1}$}\altaffiliation[]{These authors jointly supervised the study.}
\affiliation{
$^1$ Department of Mathematics \& Physics - University of Campania "Luigi Vanvitelli" - Caserta, Italy \\
$^2$  Department of Informatics, Bioengineering, Robotics, and Systems Engineering (DIBRIS), University of Genova, Genova, Italy \\
$^3$  Neurofacility, Istituto Italiano di Tecnologia (IIT), 16100 Genova, Italy\\
$^4$ Department of Engineering - University of Campania "Luigi Vanvitelli" - Aversa, Italy \\
$^5$ RCCS Ospedale Policlinico San Martino, Largo Rosanna Benzi 10, 16132, Genova, Italy \\
$^6$ Inter-University Center for the Promotion of the 3Rs Principles in Teaching \& Research (Centro 3R), Genova, Italy \\
$^7$ National Institute for Nuclear Physics (INFN), Genova, Italy}
\maketitle

\section{Analysis on individual cultures}

This Supplemental Material presents the analysis of individual experimental samples of the human iPSC-derived neuronal cultures described in the manuscript.
All methodological details, including spike detection, avalanche definition, and temporal binning procedures, are identical to those reported in Sec.~\ref{sec:Material_Method} of the manuscript.
Here, we show the sample-resolved avalanche statistics for the two experimental conditions, i.e. purely excitatory ($100$E) and physiological ($75$E:$25$I) networks, to illustrate the reproducibility and variability of the results.
Each figure shows the results of all cultures within a given condition for a specific observable: avalanche size and duration distributions, conditional expectation of size versus duration, inter-event time distributions, and conditional probability analysis.

\subsection{Static avalanche statistics: Size and Duration distributions}
\begin{figure*}[ht!]
    \centering
    \includegraphics[width=1.0\textwidth]{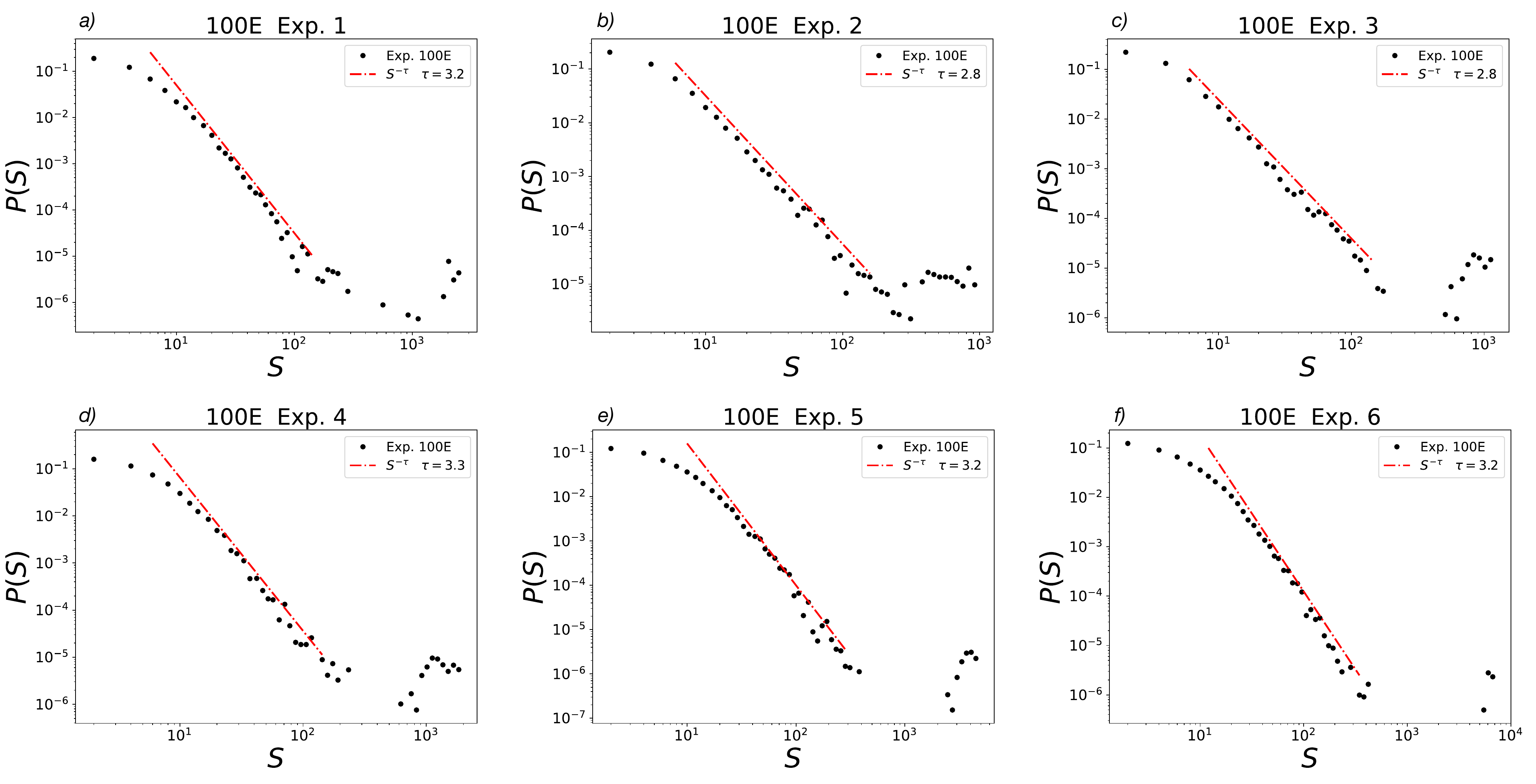}
    \label{fig:size_distribution_samples_100E}
    \caption{Avalanche size distributions with superimposed theoretical power-law behavior (dashed-dotted line) for experiments on purely excitatory networks ($100$E). The different panels correspond to the samples analyzed in the study. Panels (a-d-e-f) and (b-c) refer to the group with large and small exponents for avalanches size distributions, respectively. }
\end{figure*}

\begin{figure*}[ht!]
    \centering
    \includegraphics[width=1.0\textwidth]{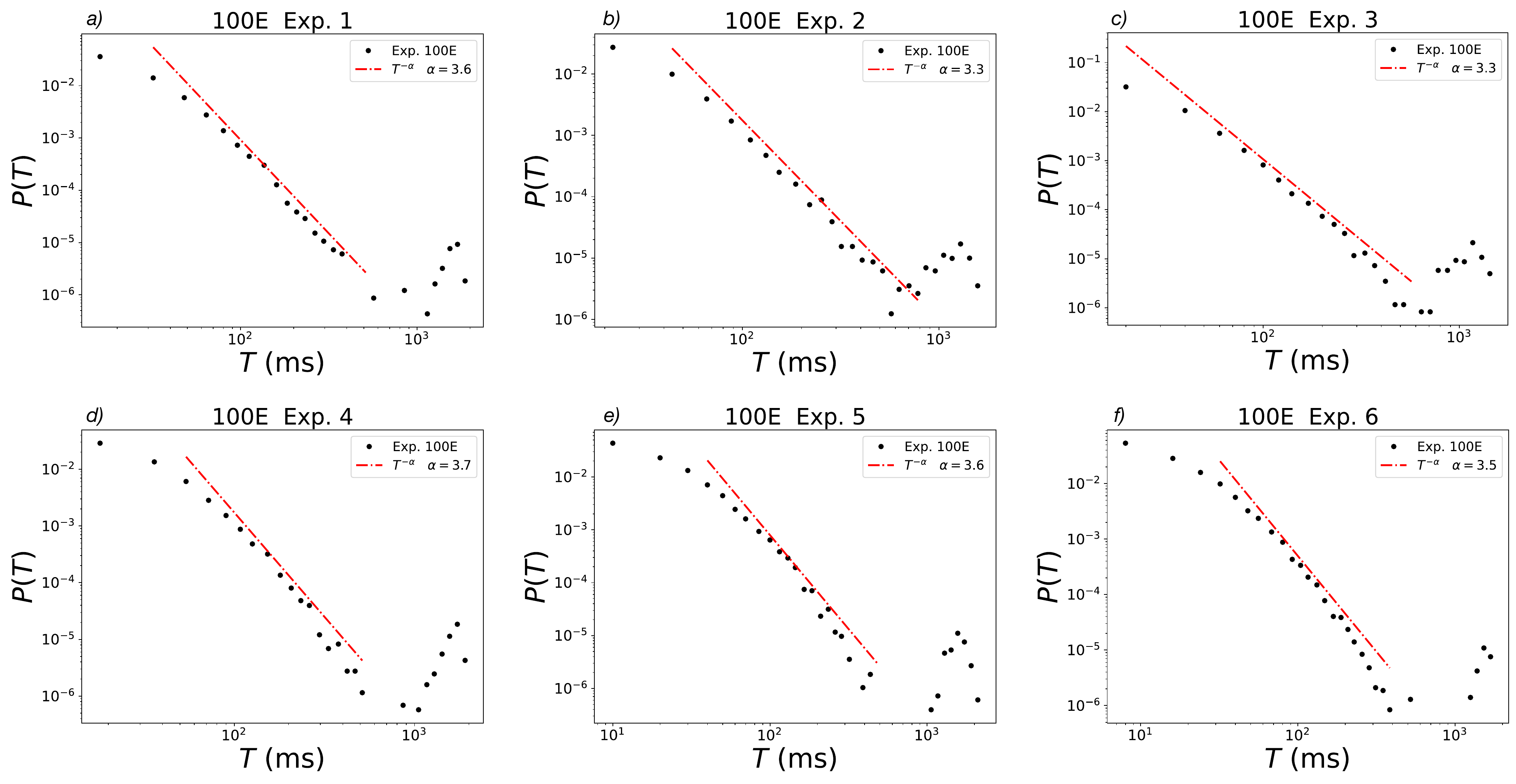}
    \caption{Avalanche duration distributions with superimposed theoretical power-law behavior (dashed-dotted line) for experiments on purely excitatory networks ($100$E). The different panels correspond to the samples analyzed in the study. Panels (a-d-e-f) and (b-c) refer to the group with large and small exponents for avalanches duration distributions, respectively. }
\label{fig:duration_distribution_samples_100E}
\end{figure*}

\begin{figure*}[ht!]
    \centering
    \includegraphics[width=1.0\textwidth]{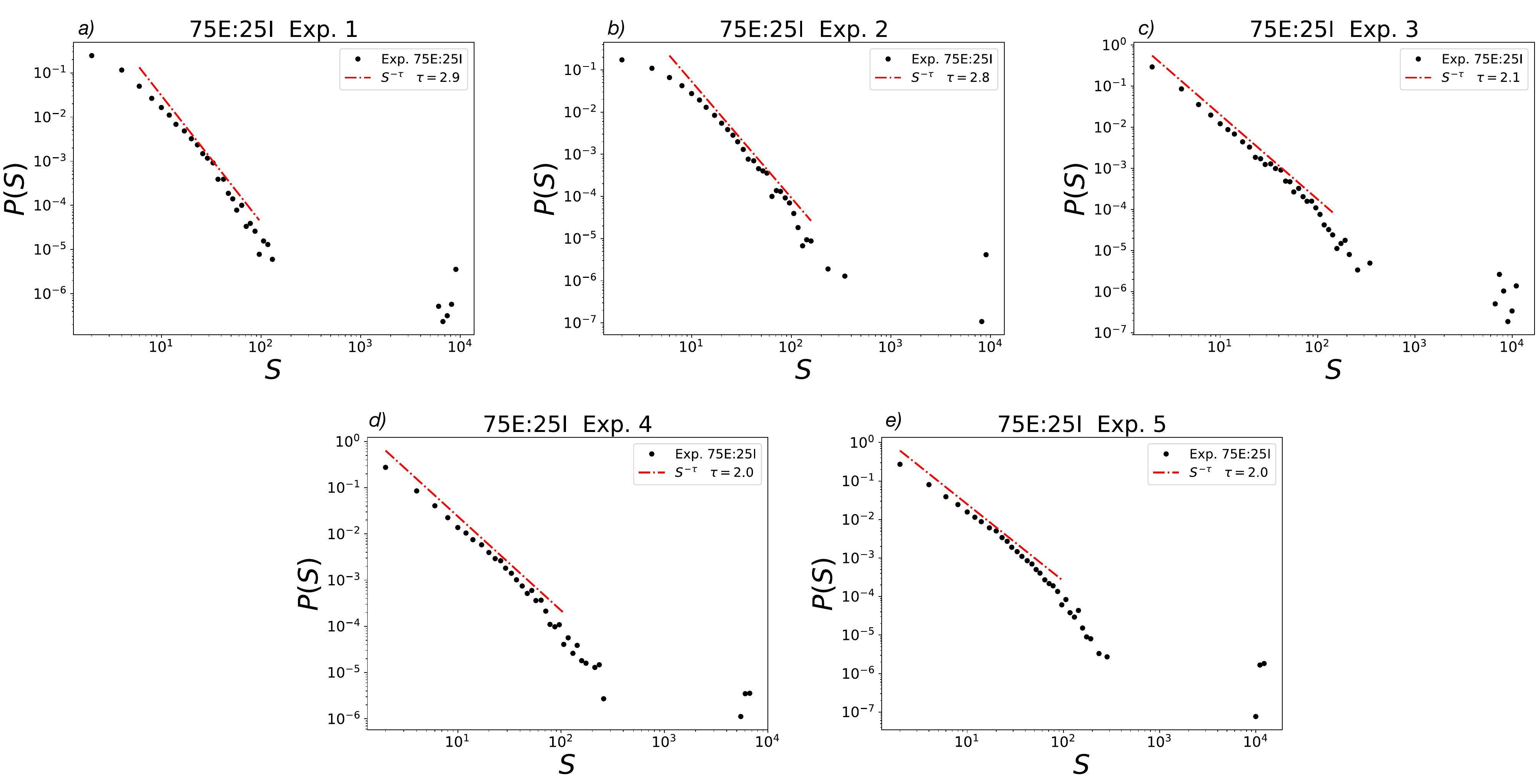}
    \caption{Avalanche size distributions with superimposed theoretical power-law behavior (dashed-dotted line) for experiments on physiological networks (75E:25I). The different panels correspond to the samples analyzed in the study. Panels (a-b) and (c-d-e) refer to the group with large and small exponents for avalanches size distributions, respectively. }
\label{fig:size_distribution_samples_75E25I}
\end{figure*}

\begin{figure*}[ht!]
    \centering
    \includegraphics[width=1.0\textwidth]{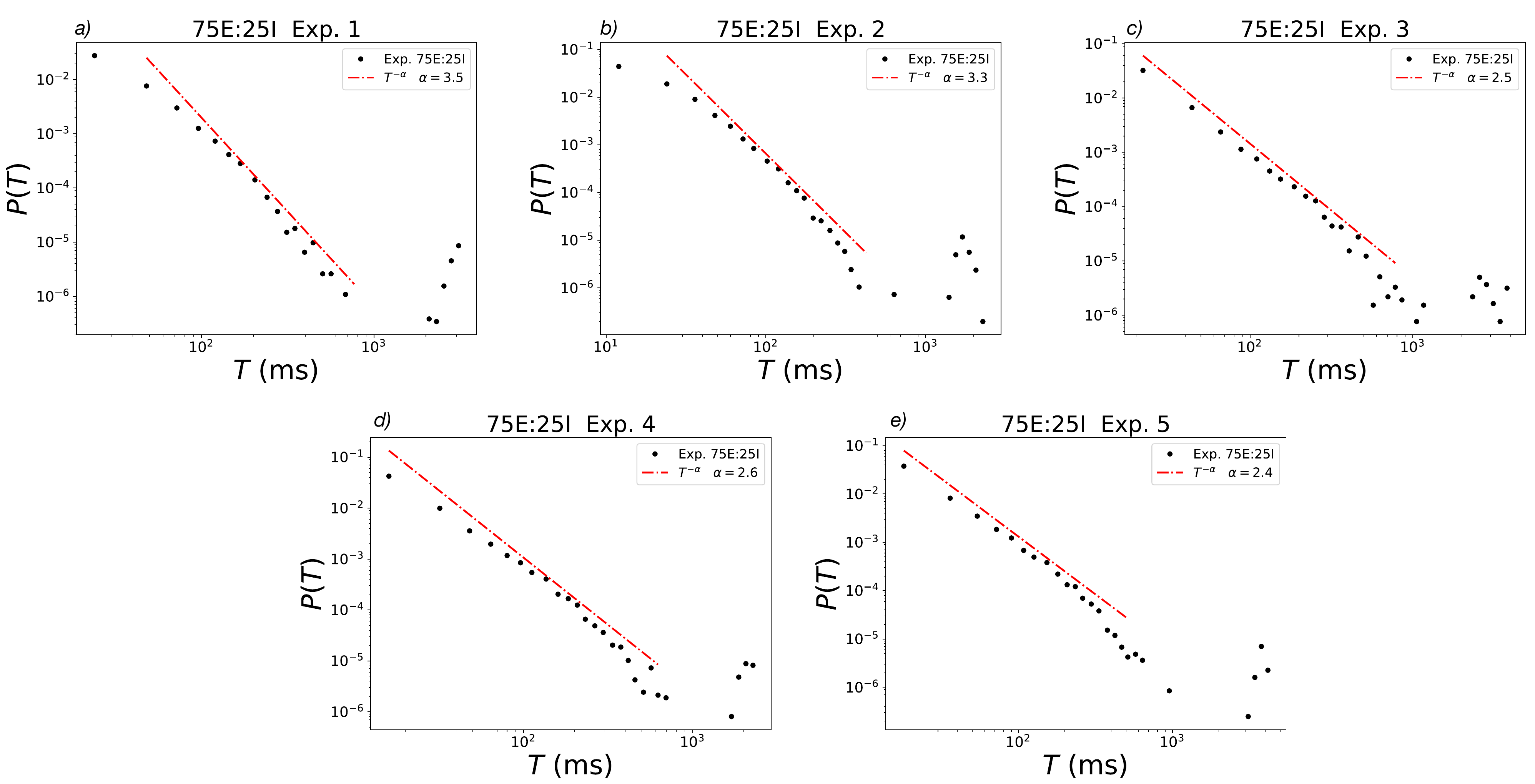}
    \caption{Avalanche duration distributions with superimposed theoretical power-law behavior (dashed-dotted line) for experiments on physiological networks (75E:25I). The different panels correspond to the samples analyzed in the study. Panels (a-b) and (c-d-e) refer to the group with large and small exponents for avalanche duration distributions, respectively. }
\label{fig:duration_distribution_samples_75E25I}
\end{figure*}

In this section we present the analysis on avalanche size and duration distributions for 100E (Figs.~\ref{fig:size_distribution_samples_100E}-\ref{fig:duration_distribution_samples_100E}) and 75E:25I (Figs.~\ref{fig:size_distribution_samples_75E25I}-\ref{fig:duration_distribution_samples_75E25I}) configurations. In all this figures, dashed lines correspond to theoretical power-law behavior whose exponents have been obtained through a fitting procedure in log-log space with the function $f(x)=ax+b$. As explained in Sec.~\ref{sec:Material_Method}, the errors have been estimated by varying the window over which the fit is performed. 
The results of the fitting procedure are summarized in Tab.~\ref{tab:results_individual_100E} of the manuscript.

\subsection{Relation between Fano factor and exponent $\tau$}
\begin{figure*}[ht!]
    \centering
    \includegraphics[width=1.0\textwidth]{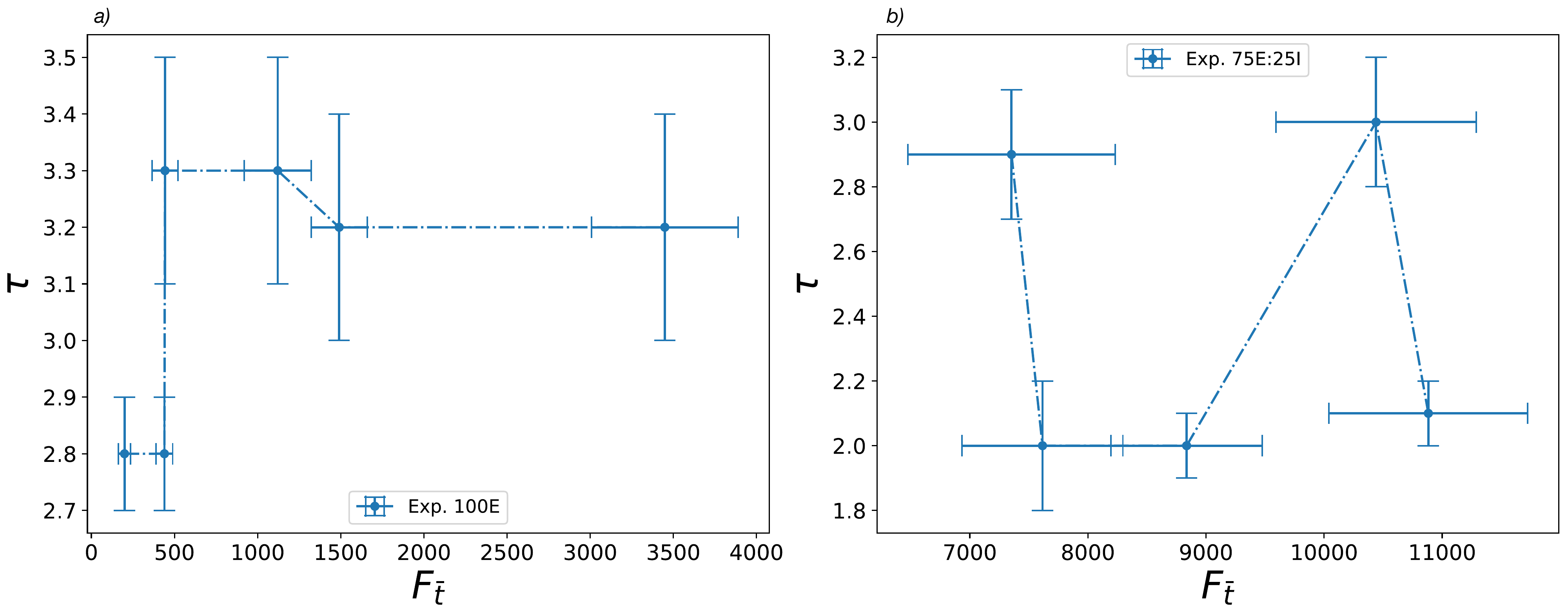}
    \caption{Exponent $\tau$ of avalanche size distribution as a function of the Fano Factor computed at $1$s $FF$. Left and right panels correspond to experiments on purely excitatory (100E) or physiological networks (75E:25I), respectively.}
\label{fig:Fano_Factor_tau}
\end{figure*}

The Fano factor $F_t$ is a measure of the dispersion of a point process with respect to the Poisson process. For a generic point process, let $N_t$ be the number of events observed in a time-window of length $t$. Then, $F_t$ is defined as the ratio of the variance $\langle N_t^2\rangle -\langle N_t\rangle^2$ to the mean $\langle N_t \rangle$, i.e. 
\begin{equation}
    F_t = \frac{\langle N_t^2\rangle -\langle N_t\rangle^2}{\langle N_t \rangle}\,.
    \label{eq:fano-factor}
\end{equation}
For a Poisson process, $\langle N_t^2\rangle -\langle N_t\rangle^2=\langle N_t \rangle$ and $F_t=1$ for every choice of $t$.
Thus, $F_t>1$ ($F_t<1$) indicates overdispersion (underdispersion) with respect to a Poisson process. 
This quantity is widely employed for the following reason: it provides a non-trivial relation between the distribution of the inter-spike times $p(\mathtt{T})$ and the distribution $p(N_t)$ of the number of spikes $N_t$ in a given time interval of duration $t$. In particular, defining the coefficient of variation $c_v$ as
\begin{equation}
    c_v = \frac{\langle \mathtt{T}^2\rangle-\langle \mathtt{T}\rangle^2}{\langle \mathtt{T}\rangle^2}\,,
\end{equation}
then
\begin{align}
    &\lim_{t\to 0} F_t = 1\,,\nonumber\\
    &\lim_{t\to\infty} F_t = c_v\,.
\end{align}
Here, we use the Fano Factor $F_t$ to quantify the sample-to-sample variability observed in experiments. In particular, we consider a time-window of length $\bar{t}=1$s, and we estimate $F_{\bar{t}}$ using Eq.~\eqref{eq:fano-factor}. The value $\bar{t}=1$s has been chosen as large as possible but still ensuring robust enough statistical predictions.   We then search for a relation between the variability of the exponents and that measured by $F_{\bar{t}}$. To estimate an error on $F_{\bar{t}}$, we divide each recording into $10$ batches each lasting $90$s, we compute $F_{\bar{t}}$ on each batch and finally we consider the mean and the standard deviation of these outcomes. Fig.~\ref{fig:Fano_Factor_tau} shows the exponent $\tau$ of avalanche size distributions for all cultures in both experimental conditions (100E and 75E:25I). For 100E configurations, cultures with larger exponent $\tau$ also exhibit larger value of $F_{\bar{t}}$ (see Fig.~\ref{fig:Fano_Factor_tau} (a)). Note that almost all samples have well separated values of $F_{\bar{t}}$, and the trend can be considered reliable. For physiological configurations, conversely, there is not a trend between avalanche exponents and Fano factor $F_{\bar{t}}$ since all the values of $F_{\bar{t}}$ are comparable within the error bars (see Fig.~\ref{fig:Fano_Factor_tau} (b)). Despite this, based on the results on purely excitatory configurations, we interpret the variability as a relevant feature of the experimental samples.

\subsection{Static avalanche statistics: Scaling relation for $\langle S \rangle_{S|T}$}
\begin{figure*}[ht!]
    \centering
    \includegraphics[width=1.0\textwidth]{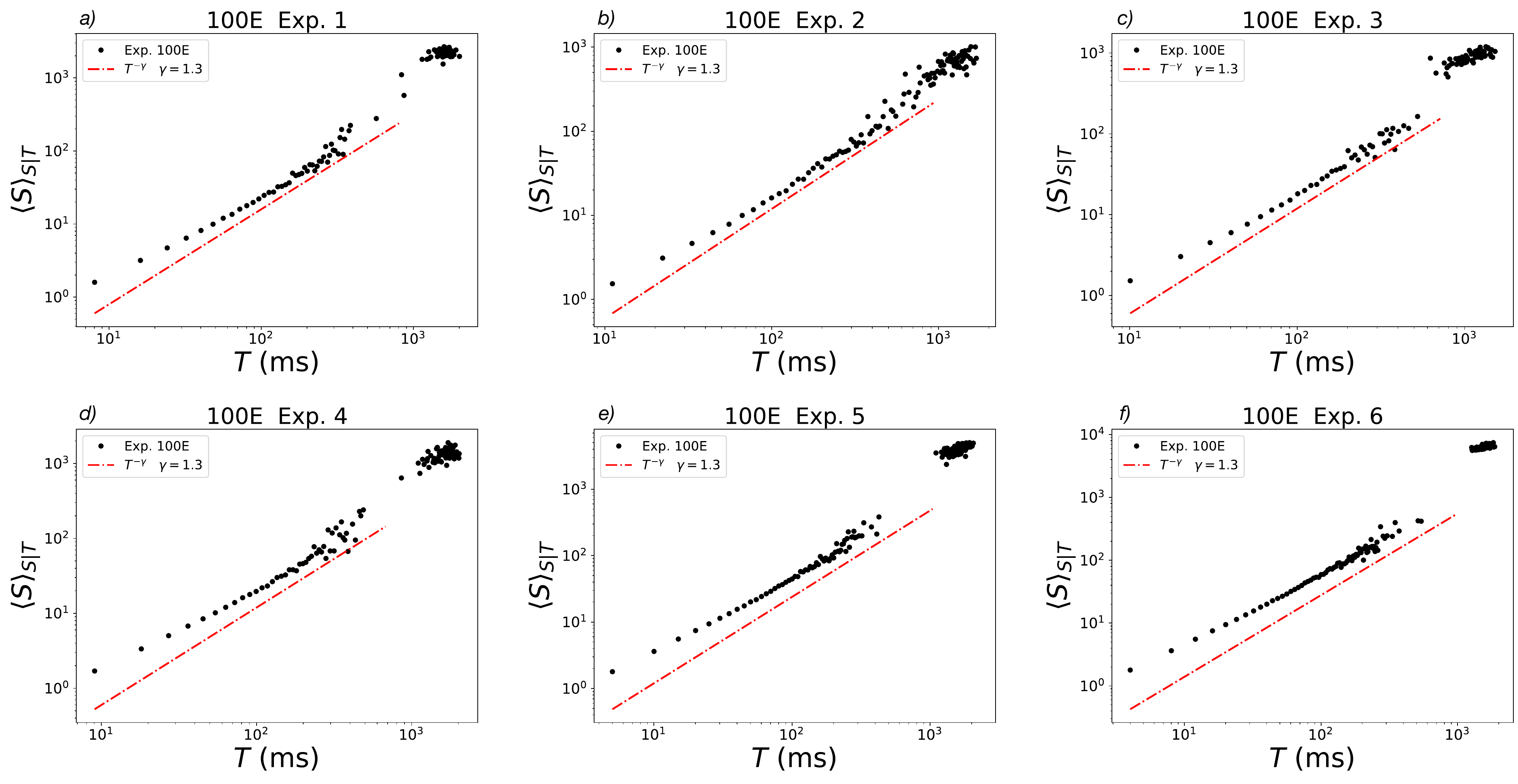}
    \caption{Conditional expectation of avalanche size $\langle S \rangle_{S|T}$ with superimposed theoretical power-law behavior (dashed-dotted line) for experiments on purely excitatory networks (100E). For the exponent $\gamma$ the average values reported in the last column of Tab.~\ref{tab:average_exponent} have been used. The different panels correspond to the samples analyzed in the study. Panels (a-d-e-f) and (b-c) refer to the group with large and small exponents for avalanches size and duration distributions, respectively. }
\label{fig:S-T_distribution_samples_100E}
\end{figure*}

\begin{figure*}[ht!]
    \centering
    \includegraphics[width=1.0\textwidth]{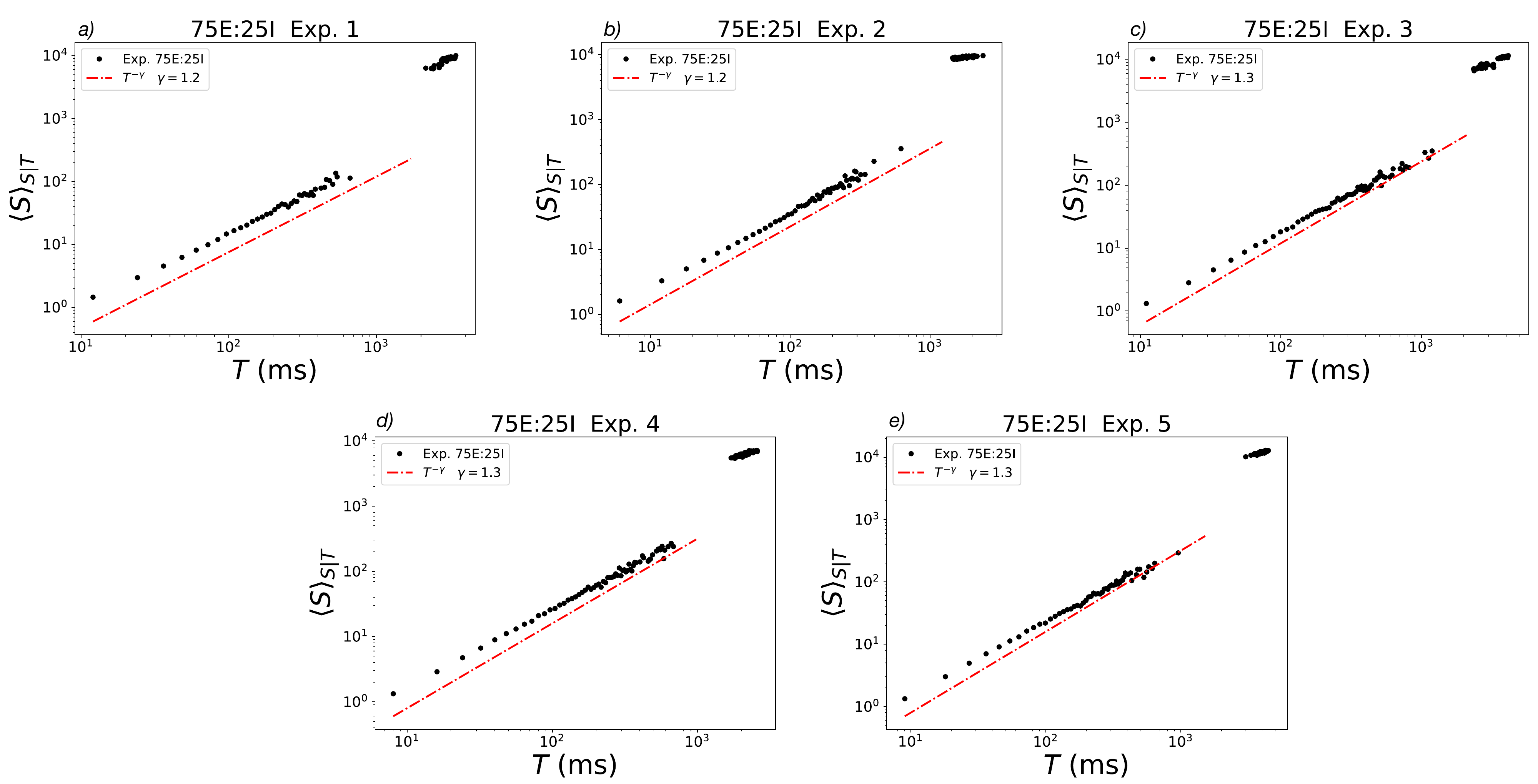}
    \caption{Conditional expectation of avalanche size $\langle S \rangle_{S|T}$ with superimposed theoretical power-law behavior (dashed-dotted line) for experiments on physiological networks (75E:25I). For the exponent $\gamma$ the average values reported in the last column of Tab.~\ref{tab:average_exponent} have been used. The different panels correspond to the samples analyzed in the study. Panels (a-b) and (c-d-e) refer to the group with large and small exponents for avalanche size and duration distributions, respectively. }
\label{fig:S-T_samples_75E25I}
\end{figure*}

Here, we show the analysis on conditional expectation $\langle S \rangle_{S|T}$ performed on individual cultures, that can be found in Figs.~\ref{fig:S-T_distribution_samples_100E}-\ref{fig:S-T_samples_75E25I} for 100E and 75E:25I configurations, respectively. It can be noted that for all cultures in both experimental conditions, the exponent $\gamma$ takes values between $[1.2,1.3]$, coherently with the analysis on the average over different groups reported in Tab.~\ref{tab:average_exponent}. The theoretical predictions appearing in Figs.~\ref{fig:S-T_distribution_samples_100E}-\ref{fig:S-T_samples_75E25I} correspond to the values of $\gamma$ reported in the last column of Tab.~\ref{tab:average_exponent}, that is those estimated on the average distributions through Eq.~\eqref{eq:hyperscaling_relation}. 

\subsection{Inter-time distribution}

\begin{figure*}[ht!]
    \centering
    \includegraphics[width=1.0\textwidth]{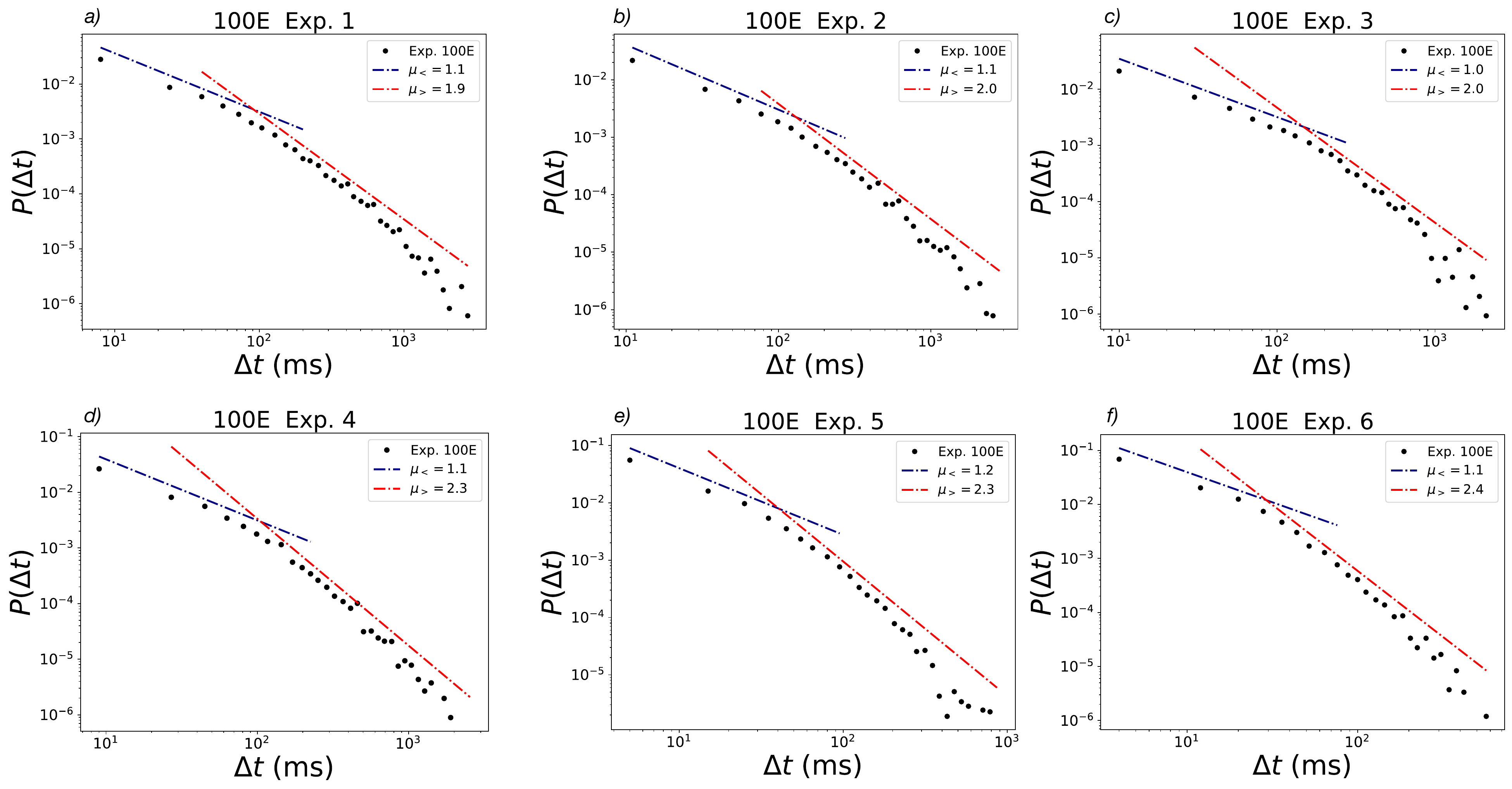}
    \caption{Inter-time distributions with superimposed theoretical power-law behavior (dashed-dotted lines) for experiments on purely excitatory networks (100E). The crossover time can be estimated from the intersection between the two power-law regimes. The different panels correspond to the samples analyzed in the study. Panels (a-d-e-f) and (b-c) refer to the group with large and small exponents for avalanches size and duration distributions, respectively. }
\label{fig:inter-time_distribution_samples_100E}
\end{figure*}

\begin{figure*}[ht!]
    \centering
    \includegraphics[width=1.0\textwidth]{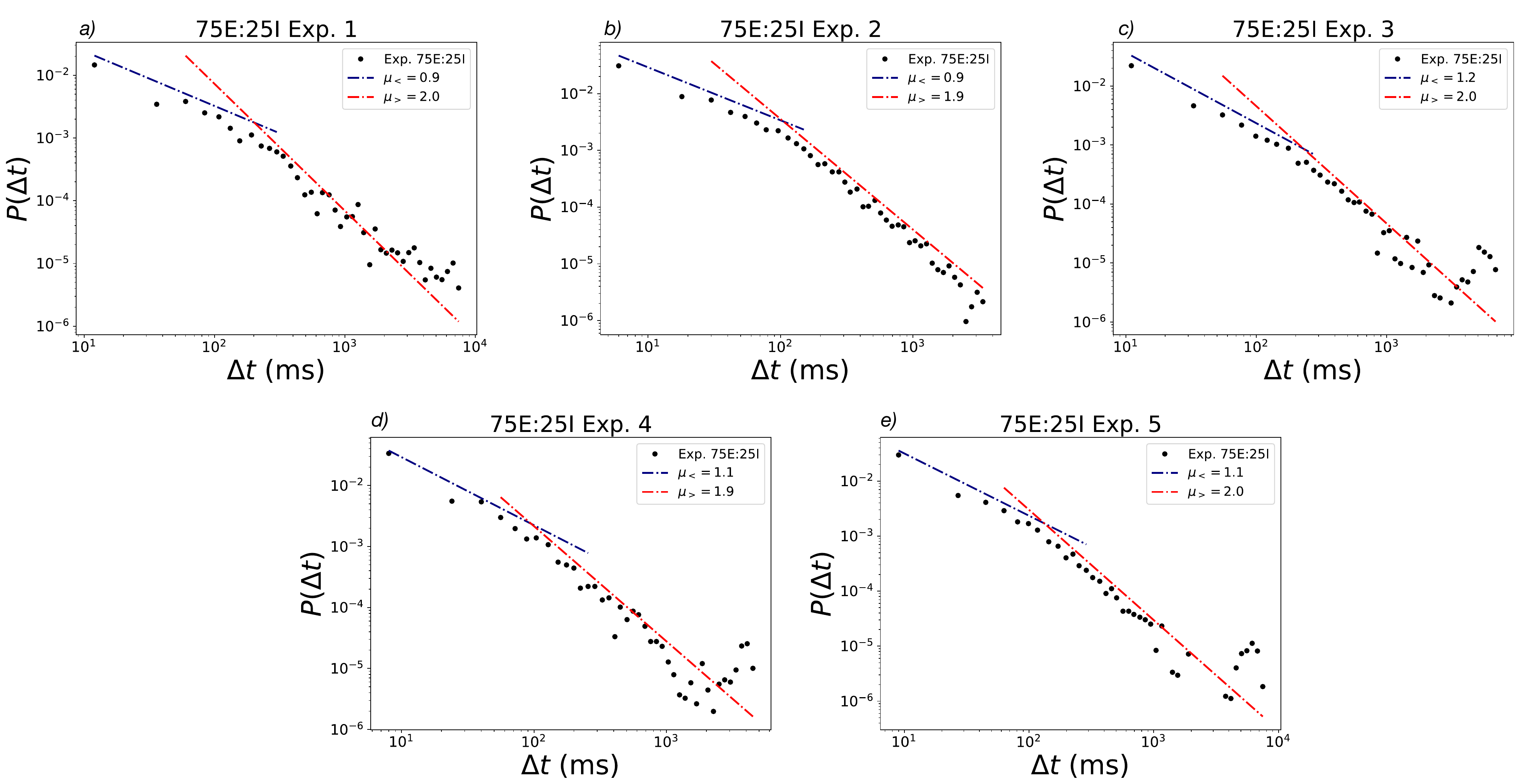}
    \caption{Inter-time distributions with superimposed theoretical power-law behavior (dashed-dotted lines) for experiments on physiological networks (75E:25I). The crossover time can be estimated from the intersection between the two power-law regimes. The different panels correspond to the samples analyzed in the study. Panels (a-b) and (c-d-e) refer to the group with large and small exponents for avalanches size and duration distributions, respectively. }
\label{fig:inter-time_distribution_samples_75E25I}
\end{figure*}

In this section, we show the analysis for inter-time distributions performed on individual cultures (see Fig.~\ref{fig:inter-time_distribution_samples_100E} and Fig.~\ref{fig:inter-time_distribution_samples_75E25I} for the analysis on purely excitatory and physiological networks, respectively). The values of the thresholds $S^\star$ and the results of the fitting procedure are those reported in Tab.~\ref{tab:inter-time_individual_75-25}. The crossover-times $\mathcal{T}_I$ are estimated as the values at which the two theoretical power-law behaviors intersect, corresponding to the results reported in Tab.~\ref{tab:crossover_individual_100}. 

\subsection{Conditional Probabilities}
\begin{figure*}[ht!]
    \centering
    \includegraphics[width=1.0\textwidth]{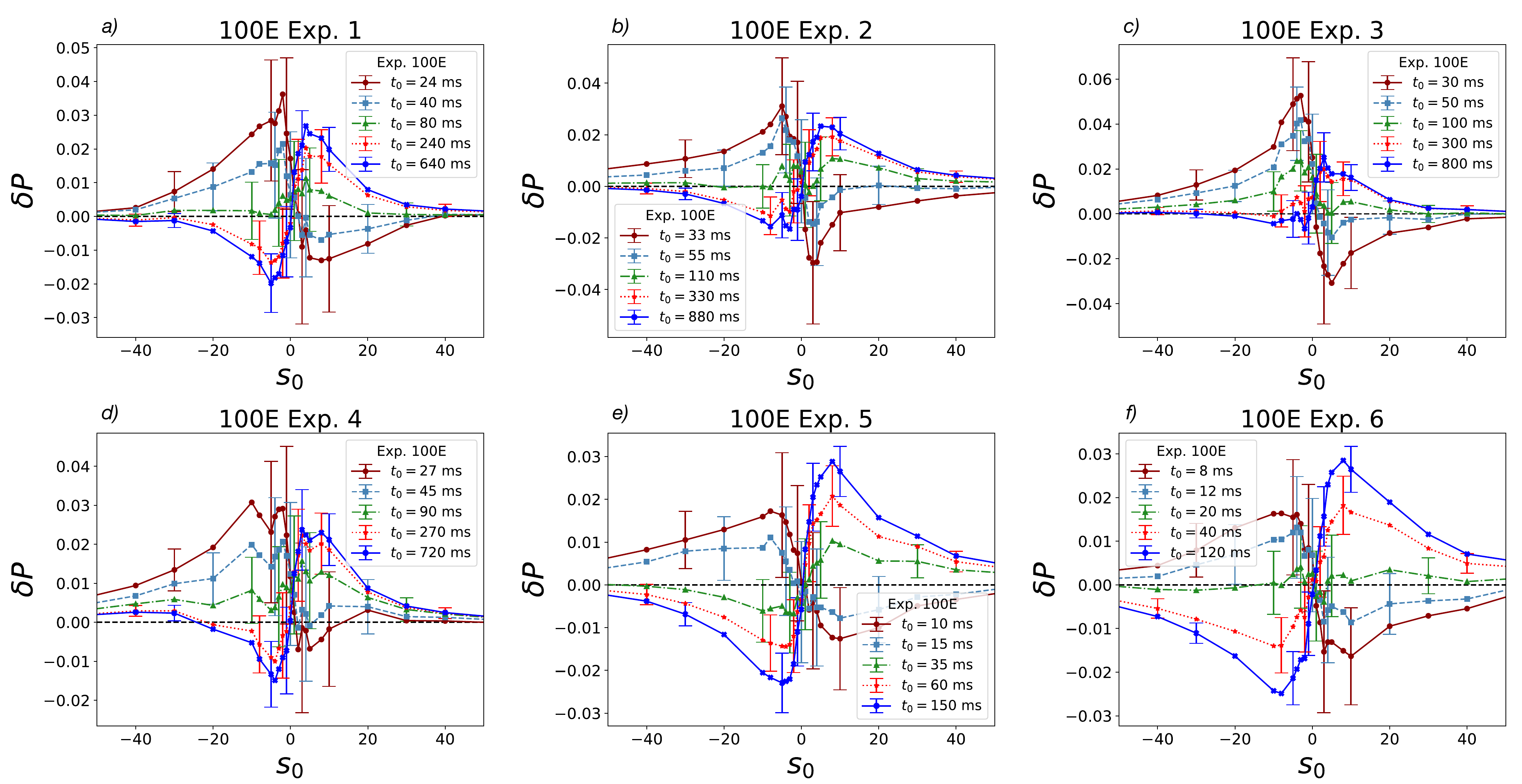}
    \caption{Difference $\delta P$ between conditional probabilities of observing a difference between subsequent avalanche sizes smaller than a given value $s_0$ conditioned on the fact that their inter-time is lower than $t_0$ for experiments on purely excitatory networks (100E). The crossover time can be estimated as the time when the switching between amplification and attenuation regime occurs. Panels (a-d-e-f) and (b-c) refer to the group with large and small exponents for avalanches size and duration distributions, respectively.}
\label{fig:conditional_probabilities_samples_100E}
\end{figure*}

\begin{figure*}[ht!]
    \centering
    \includegraphics[width=1.0\textwidth]{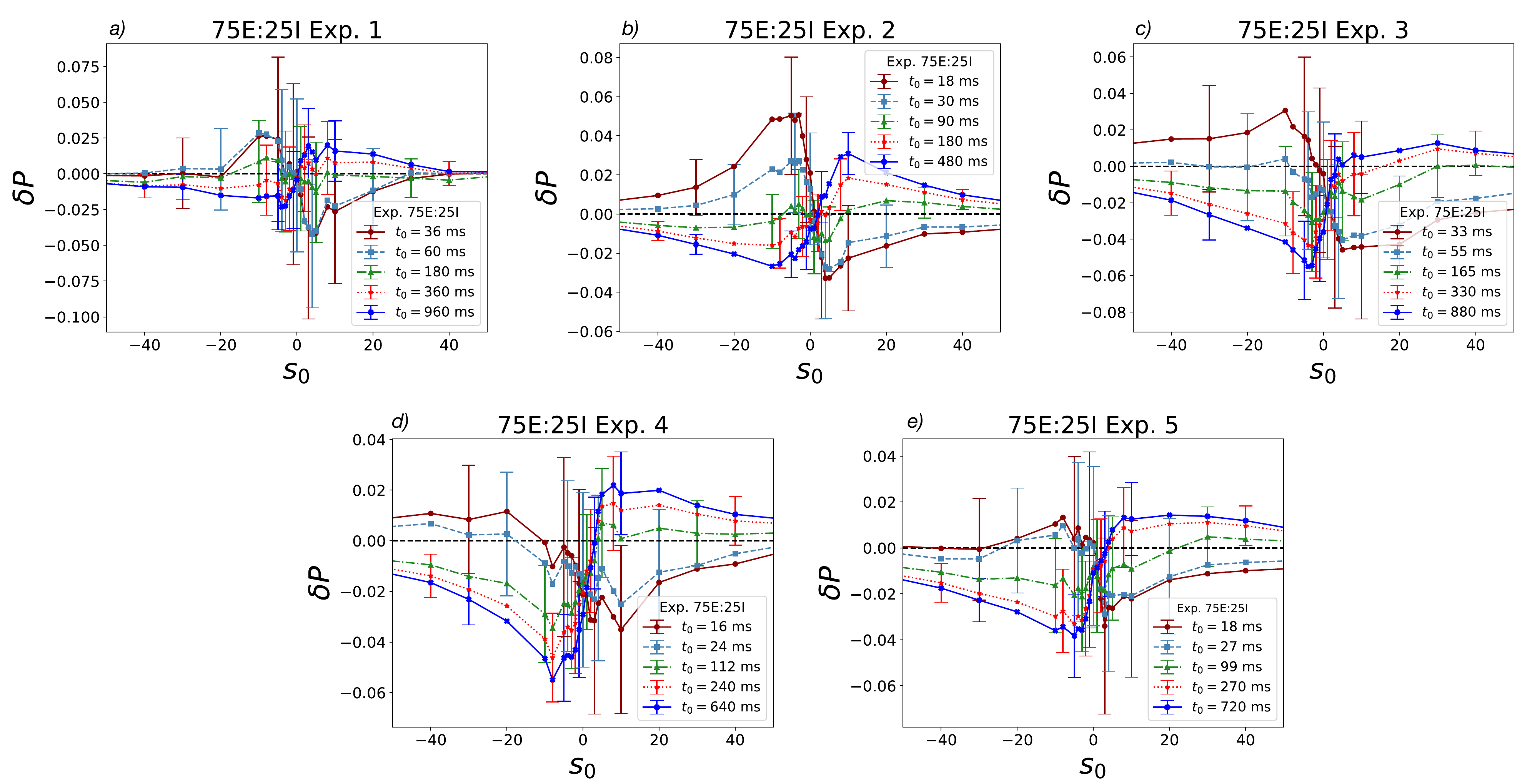}
    \caption{Difference $\delta P$ between conditional probabilities of observing a difference between subsequent avalanche sizes smaller than a given value $s_0$ conditioned on the fact that their inter-time is lower than $t_0$ for experiments on physiological networks (75E:25I). The crossover time can be estimated as the time when the switching between amplification and attenuation regime occurs. Panels (a-b) and (c-d-e) refer to the group with large and small exponents for avalanches size and duration distributions, respectively.}
\label{fig:conditional_probabilities_samples_75E25I}
\end{figure*}

As mentioned in the manuscript, the crossover between amplification and attenuation regimes is manifested not only when experiments belonging to the same group are analyzed merging their statistics but also at the level of single cultures. This is evident by looking at Figs.~\ref{fig:conditional_probabilities_samples_100E}-\ref{fig:conditional_probabilities_samples_75E25I} showing the conditional probability differences for 100E and 75E:25I configurations. From these figures, we also estimate the crossover-times $\mathcal{T}_C$ reported in Tab.~\ref{tab:crossover_individual_100}, as the values of $t_0$ such that the curves $\delta P(s_0,t_0)$ are almost flat.  

\section{Matching model-experiment timescales}
In this section we present the approach adopted for determining the physical timescale in numerical simulations. As briefly explained in the manuscript, this timescale is determined by matching the numerical time-bin $t_b$ with the average experimental time-bin. The average experimental time-bin is computed by averaging the values of $t_b$ over the experiments with the same avalanche distribution exponents $\tau$ and $\alpha$, and the results are presented in Tab.~\ref{tab:average_time_bin}. The error associated to the average $t_b$ is estimated as half the maximum variation between experiments belonging to the same group. Note that the relative error on the average $t_b$ for purely excitatory samples with larger exponents exceeds the $35\%$. Therefore, some disagreement between numerical experimental results can be expected in this case. Expressing the numerical time-bin $t_b$ in ms allows the comparison between numerical and experimental crossover-times presented in the manuscript. In particular, by multiplying the adimensional crossover-times reported in Tab.~\ref{tab:adimensional-crossover-time} with the corresponding dimensional $t_b$, we obtain the results presented in Tab.~\ref{tab:comparison-time_bin_fit} of the manuscript. 

\begin{table}[h!]
\centering
\begin{tabular}{ |c|c| } 
 \hline
Data &  Exp. $t_b$ \\[0.5ex]
 \hline
Exp $100$E  ($\tau=2.8$, $\alpha=3.3$)& $10.5\pm0.5$ \\
\hline
Exp $100$E ($\tau=3.1$, $\alpha=3.7$)& $6.5\pm2.5$\\
\hline
Exp $75$E$25$I ($\tau=2.1$, $\alpha=2.5$) & $9.3\pm1.5$ \\
\hline
Exp $75$E$25$I ($\tau=2.9$, $\alpha=3.3$) & $9.0\pm3.0$\\
\hline
\end{tabular}
\caption{Average experimental time-bin $t_b$ for the groups with different avalanche distribution exponents $\tau$ and $\alpha$.} 
\label{tab:average_time_bin}
\end{table}

\begin{table}[h!]
\centering
\begin{tabular}{ |c|c|c| } 
 \hline
Data &  $\mathcal{T}_I$ ($t_b$) & $\mathcal{T}_C$ ($t_b$)\\[0.5ex]
 \hline
$p=10^{-5}$ $100$E & $14\pm8$ & $7\pm5$ \\
\hline
$p=10^{-4}$ $100$E & $19\pm7$ & $9\pm5$ \\
\hline
$p=10^{-4}$ $75$E$25$I & $13\pm 10$ & $8\pm5$ \\
\hline
$p=5\cdot10^{-4}$ $75$E$25$I & $19\pm 8$ & $8\pm5$ \\
\hline
\end{tabular}
\caption{Adimensional crossover times for numerical simulations extrapolated from inter-time distributions $\mathcal{T}_I$ and from conditional probability differences $\mathcal{T}_C$.} 
\label{tab:adimensional-crossover-time}
\end{table}